\documentstyle[prb,aps,epsf,array,multicol,rotate]{revtex}
\newcommand{\lrule}{ \end{multicols} \noindent
  \rule{0.5\textwidth}{0.1mm}\rule{0.1mm}{3pt}\newline }
\newcommand{\rrule}{ \noindent \parbox{\textwidth}{
  \hfill\rule[-3pt]{0.1mm}{3pt}\rule{0.5\textwidth}{0.1mm}}
  \begin{multicols}{2}\noindent }

\def\Rv{{\mbox{\boldmath$R$}}}

\def\gl{{g_L}}
\def\gr{{g_R}}
\def\fl{{f_L}}
\def\fr{{f_R}}
\def\tl{{\tilde f_L}}
\def\tr{{\tilde f_R}}

\def\cg{\check{g}}
\def\cgia{\check{g}_{\infty,i}}
\def\cgiR{\check{g}_{\infty,R}}
\def\cgiL{\check{g}_{\infty,L}}
\def\ce{\check{\epsilon}}
\def\cD{\check{\Delta}}
\def\ct{\check{t}}
\def\cv{\check{v}}

\def\ot{\mbox{$\scriptscriptstyle \otimes$}}

\def\vf{{v_{\!{\scriptscriptstyle F}}}}

\def\pfhv{\hat{\mbox{\boldmath$p$}}_{\!\!{\scriptscriptstyle F}}}
\def\ppfhv{\hat{\mbox{\boldmath$p$}}^\prime_{\!\!{\scriptscriptstyle F}}}

\def\sfhv{\hat{\mbox{\boldmath${\scriptstyle p}$}}_{\!\!{\scriptscriptstyle F}}}
\def\nhv{\hat{\mbox{\boldmath$n$}}}
\def\vfv{{\mbox{\boldmath$v$}}_{\!\!{\scriptscriptstyle F}}}

\def\sRv{{\mbox{\boldmath \mbox{$\scriptscriptstyle R$}}}}
\def\qcgrad{{i\, \vfv\!\!\cdot\!\!\nabla_{\!\!\sRv}\, }}

\begin{document}
\title{Quasiclassical description of transport through superconducting contacts}
\author{J.~C. Cuevas}
\address{
Institut f\"ur Theoretische Festk\"orperphysik, Universit\"at 
Karlsruhe, D-76128 Karlsruhe, Germany\\
}
\author{M. Fogelstr\"om}
\address{
Institute of Theoretical Physics, Chalmers University of 
Technology and G\"oteborgs University, S-41296 G\"oteborg, Sweden\\
Institut f\"ur Theoretische Festk\"orperphysik, Universit\"at 
Karlsruhe, D-76128 Karlsruhe, Germany
}
\date{\today}
\maketitle
\begin{abstract}
We present a theoretical study of transport properties through superconducting contacts 
based on a new formulation of boundary conditions that mimics interfaces for the 
quasiclassical theory of superconductivity. These boundary conditions are based on a description 
of an interface in terms of a simple Hamiltonian. We show how this Hamiltonian description is 
incorporated into quasiclassical theory via a T-matrix equation by integrating out irrelevant 
energy scales right at the onset. The resulting boundary conditions reproduce results obtained by 
conventional quasiclassical boundary conditions, or by boundary conditions based on the scattering 
approach. This formalism is well suited for the analysis of magnetically active interfaces as well 
as for calculating time-dependent properties such as the current-voltage characteristics or as current 
fluctuations in junctions with arbitrary transmission and bias voltage. This approach is illustrated 
with the calculation of Josephson currents through a variety of superconducting junctions ranging 
from conventional to d-wave superconductors, and to the analysis of supercurrent through a 
ferromagnetic nanoparticle. The calculation of the current-voltage characteristics and of noise 
is applied to the case of a contact between two d-wave superconductors. In particular, we discuss 
the use of shot noise for the measurement of charge transferred in a multiple Andreev reflection in 
d-wave superconductors.

\end{abstract}


%===============================================================================================
\begin{multicols}{2}
\section{Introduction}

At the end of the 1960's several authors demonstrated that the complete standard theory
of superconductivity, both in equilibrium and out of equilibrium, can be formulated in
terms of a quasiclassical transport equation \cite{Quasiclassics1968}. This theory, known as 
{\it quasiclassical theory of superconductivity}, combines Landau's semiclassical transport equations 
for quasiparticles with the concept of pairing and particle-hole coherence that are the basis
of the BCS theory. The quasiclassical theory provides a full description of superconducting 
phenomena ranging from inhomogeneous superconductors to superconducting phenomena far from
equilibrium. In its traditional form, the quasiclassical theory of superconductivity 
is restricted to the description of weak perturbations of a superconductor: the external perturbations
(magnetic field, variations in the chemical potential, etc) should be weak ($V \ll E_F$), where $E_F$ is 
the Fermi energy, of long wavelength ($q \gg k_F^{-1}$), where $k_F^{-1}$ is the Fermi wavelength, and of low 
frequency ($\hbar \omega \ll E_F$) \cite{Serene1983}. Although this limit covers many situations of
interest in superconductivity, some important cases fall outside the range of
validity of the usual scheme of the quasiclassical theory: strong impurities,
walls, interfaces, etc. In the case of the analysis of the transport properties
of superconducting contacts, the quasiclassical theory is complemented
by appropriate boundary conditions, the so-called Zaitsev boundary conditions
\cite{Zaitsev1984}, or by its generalization to magnetically active interfaces by Millis,
Rainer and Sauls \cite{Millis1988}. Although these boundary conditions provide a formal solution
of the problem of a strong perturbation due to interfaces, their highly non-linear
form introduces many problems, e.g. these boundary conditions
have spurious solutions which require special care, in particular in numerical
implementations. At this point, we should mention that the recent progress in the solution
of these boundary conditions make them much more tractable \cite{Eschrig2000,Shelankov2000,Fogelstrom2000}.

In the field of electronic transport through superconducting junctions there are many
basic situations in which a complete understanding is still lacking. This is partially due to the 
difficulty of applying the quasiclassical theory because of the lack of simple and manageable 
boundary conditions. 
Most of these situations are related to either the current-voltage (I-V) characteristics
of contacts between two superconductors with arbitrary transmission, or to situations in which there are 
magnetically active interfaces giving rise to spin-dependent transport. For instance, in the context of
high temperature superconductors (HTS), which by now are believed to be d-wave superconductors 
\cite{Harlingen95}, we can mention N-I-S tunnel 
junctions\cite{Covington1997,Aprili1998,Greene2000,Covington2000},
grain boundary junctions widely used for the realization of Josephson junctions 
\cite{Alff98,Koelle99,Tsuei2000}, and YBCO S-N-S edge junctions \cite{Polturak93,Engelhardt99,Nesher99}.
In the two latter systems the description of the I-V characteristics requires an analysis of multiple 
Andreev reflections (MAR) \cite{Hurd97}, while all systems above require a careful self-consistent study of 
the superconducting state in the presence of pair-breaking, caused by bulk or surface impurities and by
the surface itself, in order to account for results obtained in experiments
\cite{Covington1997,Aprili1998,Greene2000,Covington2000} by theory
\cite{Barash1996,Poenicke99,Fogelstrom-preprints,Poenicke00}.
Other examples, which take place in the context of conventional superconductors, are the S-N-S point contacts,
where the normal region $N$ has a length comparable to the superconducting coherent length 
\cite{Takayanagi95,Scheer2000}. These systems demand from the theory a detailed analysis of the interplay 
between the proximity effect superconductivity in the normal metal and the occurrence of MAR. This 
analysis is currently in progress making use of the approach described in this work \cite{Fogelstrom2001}.

With respect to the transport through spin active interfaces 
there have been an extensive experimental effort to explore
tunneling through magnetic insulator barriers\cite{STA85,MT94},
by probing the proximity effect in S/F-structures\cite{Giroud1998,Petrashov1999},
and by constructing S/F-multilayers\cite{CR99} (see also references therein). 
Andreev reflection using ferromagnet/superconductor point contacts
have also been used to probe the spin polarization of the ferromagnet\cite{Soulen1998,Upadhyay1998}.
Finally, 
supercurrents have been reported in S/F/S junctions by Veretennikov {\it et al} \cite{V00} and by
Gandit {\it et al} \cite{Gandit2000}.
To model an S/F/S-junction one can follow one of two routes. The first route is to
assume an extension of a ferromagnetic metal, now characterized by a length and an exchange field,
separating the two superconductors.  Following this route, both critical current oscillations\cite{BBP82,RAD91} and
the effect of the exchange field on the Andreev bound states\cite{KF90,RAD99} have been studied. The limitation
here is that the approach is restricted to small exchange fields in the ferromagnet, 
i.e. fields that are comparable to the
superconducting gap. The second route, which we subscribe to, is to treat the effect of the ferromagnet as a 
boundary problem connecting the two superconducting half-spaces\cite{BKS77,DA85,Millis1988,JB95}.
Using Bogoliubov-deGennes equations and a WKB-approach for the ferromagnetic barrier, i.e. the scattering
approach\cite{JB95},
Josephson current-phase relations\cite{KTYB97} and quasiparticle tunneling\cite{KTYB99,ZV00} have been
studied for both conventional s-wave and unconventional d-wave superconductors.

Parallel to the quasiclassical treatment of interfaces, a different formalism referred to as the Hamiltonian approach
has emerged as a new interesting tool for the analysis of transport through
superconducting contacts\cite{MartinRodero1994,Cuevas1996}. This approach is based on
modeling the contact by a simple Hamiltonian in combination
with non-equilibrium Green's function techniques\cite{Keldysh1965}. 
The origin of this approach can be traced back 
to the work of Bardeen who introduced the tunnel Hamiltonian approximation in order to
describe a tunnel junction \cite{Bardeen61}. Most of the calculations of
current through superconducting contacts in the early 60's were based on this
tunnel Hamiltonian, and were restricted to the lowest order transport processes
such as the calculation of the Josephson current in a S-I-S junction
\cite{Josephson62,Ambegaokar63}. Multi-particle tunneling was first discussed
by Schrieffer and Wilkins \cite{MPT} in their \emph{multiparticle tunneling
theory} as a possible explanation for the observed sub-gap structure in
superconducting tunnel junctions. The contributions of these higher order
processes were found to be divergent, which has led to the quite extended
belief that the Hamiltonian approach is pathological except for describing the
lowest order tunneling processes. However, in a series of works, Caroli and
coworkers in the 70's \cite{Caroli71} and, more recently, Mart\'{\i}n-Rodero and collaborators
\cite{MartinRodero1994,Cuevas1996} have shown that one can get rid of all
the old pathologies of this approach by using a local representation and by summing the series of 
tunneling processes up to infinite order. In particular, within this technique a
microscopic theory of Multiple Andreev reflections has been developed \cite{Cuevas1996}.
This theory describes quantitatively the I-V characteristic of superconducting atomic-size 
contact\cite{Scheer97,Scheer98}.

In this paper we show how the Hamiltonian approach may be brought into the
quasiclassical theory by integrating out large energies right at the onset.
The resulting boundary condition reproduces results obtained by the
conventional quasiclassical Zaitsev boundary condition\cite{Zaitsev1984,Eschrig2000} or
by boundary conditions based on the scattering approach\cite{Landauer,Buttiker,BTK,Shelankov1984,Beenakker1991}. 
The boundary condition is also well suited for calculating time-dependent properties
such as the current-voltage
characteristics or the current fluctuations of S-I-S junctions. The paper is organized
in the following manner: In section II we give the energy integration of the Hamiltonian
approach and state the resulting boundary condition. In section III we show how
the current through a contact may be calculated from the boundary t-matrix. In this section
we also calculate the Josephson current resolved in energy and on trajectory for
different types of superconductors and for different types of coupling between the
two superconductors. In section IV, we discuss the boundary condition at a finite
bias applied between the superconductors. This is then applied to the case of two
coupled d-wave superconductors. Finally, in section V, current-fluctuations are
discussed and the current-current correlator is derived within our method. This is then
applied to compute the trajectory resolved current-fluctuations of two coupled d-wave superconductors.
%===============================================================================================
\section{Description of the approach}
The system of study is two semi-infinite superconducting electrodes coupled over
some type of interface barrier. Our approach is to artificially separate the problem in to 
two parts in order to pose a boundary condition for the interface.
The first part consists of calculating the Green's function of either conductor, extending to $\pm\infty$ 
respectively, in presence of a hard surface at $x=0$. To this part of the problem, the quasiclassical 
theory\cite{Quasiclassics1968} is our theory of choice. It has been shown that strong perturbations, such 
as rigid walls may be included into quasiclassical theory by means of effective boundary conditions
posed for the quasiclassical Green's function\cite{Buchholtz1979,Serene1983,Kurkijarvi1987}.
To couple the two electrodes, from now denoted left (L) and right (R), we assume a phenomenological
Hamiltonian as follows\cite{MartinRodero1994,Cuevas1996}
\begin{equation}
\hat{H}_T = \sum_{\sigma} \hat{c}^{\dagger}_{L,\sigma} v_{LR}
\hat{c}_{R\sigma} + \hat{c}^{\dagger}_{R\sigma} v_{RL} \hat{c}_{L\sigma}.
\label{hopping}
\end{equation}
The potentials $v_{LR}$ and $v_{RL}$, with $v^{\dagger}_{RL}=v_{LR}=v$, act as hopping elements connecting
the two electrodes $L$ and $R$. The perturbation given by $\hat{H}_T$ is short ranged ($\sim \lambda_F\ll\xi_o$)
and may be strong ($v\sim E_F$). The local character of $\hat{H}_T$ allows us to view it as a
single strong impurity in case of a point contact or as a line of strong impurities in case of
an extended contact between the two electrodes.
Following the work of Thuneberg and co-workers\cite{Thuneberg1981,Thuneberg1984} 
the single impurity or, in case of the line of impurities following the work of Buchholtz and Rainer\cite{Buchholtz1979},
this strong perturbation may also be incorporated into quasiclassical theory via a T-matrix equation. 
Anticipating the result for the T-matrix, the effect of the contact between the two electrodes on the 
physical quasiclassical Keldysh-Nambu matrix Green's function, or propagator,
$\cg_i$, in electrode $i=(L,R)$ enters as a source term in the 
transport equation for $\cg_i(\pfhv)$
along a trajectory $\pfhv$
\begin{eqnarray}
\qcgrad \cg_i(\pfhv)&+&\lbrack\ce_i(\pfhv)-\cD_i(\pfhv),\cg_i(\pfhv)\rbrack_{\ot}\nonumber\\
&=&\lbrack \ct_{ii}(\pfhv,\pfhv),\cgia(\pfhv) \rbrack_{\ot} \delta(\Rv-\Rv_{c}).
\label{transport}
\end{eqnarray}
Here $\Rv_{c}$ is the position of the contact and $\vfv$ is the Fermi velocity at point $\pfhv$ on the Fermi surface.
The Green's function $\cgia$ is an intermediate Green's function obtained by solving the hard wall boundary
condition of the separate electrodes, i.e. without taking the contact into account but using the self energies 
$\ce_i$ and $\cD_i$ evaluated using the physical propagator, $\cg_i$, satisfying Eq. (\ref{transport}). 

Our objective is to find the quasiclassical T-matrix, $\ct$, giving the source term in the transport equation 
(\ref{transport}) above. The starting point is a conventional many-body perturbation theory for the 
Hamiltonian $\hat{H}_T$. To proceed we artificially enlarge our Hilbert space with a "reservoir quantum number" 
$(L,R)$ and the functions entering are the matrices 
\[
\begin{array}{ll}
\tilde{\check{G}}=\left(\begin{array}{cc}\check{G}_{LL} & \check{G}_{LR}  \\
                                         \check{G}_{RL} & \check{G}_{RR} \end{array}\right) &
\tilde{\check{T}}=\left(\begin{array}{cc}\check{T}_{LL} & \check{T}_{LR} \\ 
                                         \check{T}_{RL} & \check{T}_{RR} \end{array}\right) \\*[0.5truecm]
\tilde{\check{G}}_\infty=\left(\begin{array}{cc}\check{G}_{\infty,L} & 0 \\ 
                                                                   0 & \check{G}_{\infty,R} \end{array}\right) &
\tilde{\cv}=\left(\begin{array}{cc} 0 & \cv_{LR} \\ 
                             \cv_{RL} & 0 \end{array}\right).
\end{array}
\]
The matrix elements are the usual Keldysh-Nambu matrices of non-equilibrium superconductivity\cite{Serene1983}.
Especially, the Green's functions $\check{G}_{\infty,L}$ and $\check{G}_{\infty,R}$ are the Green's
functions for the uncoupled left and right electrode. The coupling elements, $\cv_{LR,RL}$, between $L$ and 
$R$ are proportional to the unit matrix in the Keldysh space and in Nambu space adopt the form

\[
\hat{v}_{LR} = \hat{v}^{\dagger}_{RL}  = \left(\begin{array}{cc} v & 0 \\
0 & -v^{\dagger} \end{array}\right) .
\]

\noindent
With this, we write the T-matrix equation 
\begin{eqnarray}
\tilde{\check{T}}&=&\tilde{\cv}+\tilde{\cv}\circ\tilde{\check{G}}_\infty\circ\tilde{\check{T}}
\label{Tequation1}\\
                 &=&\tilde{\cv}+\tilde{\cv}\circ\tilde{\check{G}}\circ\tilde{\cv} ,
\label{Tequation2}
\end{eqnarray}
which together with the Dyson equation
\begin{eqnarray}
\tilde{\check{G}}&=&\tilde{\check{G}}_\infty
    +\tilde{\check{G}}_\infty\circ\tilde{\check{T}}\circ\tilde{\check{G}}_\infty
\label{Gequation1}\\
                 &=&\tilde{\check{G}}_\infty+\tilde{\check{G}}_\infty\circ\tilde{\cv}\circ\tilde{\check{G}} 
\label{Gequation2}
\end{eqnarray}
constitutes a closed set of equations that are to be brought into quasiclassical form. 
We have given two different ways of summing the series which correspond either
to "dressing" the perturbation (equations (\ref{Tequation1}) and (\ref{Gequation1}))
or to "dressing" the Green's function (equations (\ref{Tequation2}) and (\ref{Gequation2})). 
The two sets of equations are equivalent and two useful relations 
\begin{equation}
\tilde{\cv}\circ\tilde{\check{G}}=\tilde{\check{T}}\circ\tilde{\check{G}}_\infty \quad\mbox{and}\quad
\tilde{\check{G}}_\infty\circ\tilde{\check{T}}=\tilde{\check{G}}\circ\tilde{\cv}
\label{TGI_vGrelatons}
\end{equation}
follow directly. Here, and above, the $\circ$-product is short-hand for 
integration or summation over common arguments.
Starting from equation (\ref{Tequation2}), using equation (\ref{Gequation2}) and the second of the two relations
(\ref{TGI_vGrelatons}), it is straight forward to get the following closed set of equations, closed separately for
one and each of the components, $\check{T}_{ij}$, of the T-matrix\cite{Cuevas1996},
\begin{eqnarray}
\check{T}_{LL}&=&\cv_{LR}\circ \check{G}_{\infty,R}\circ\cv_{RL}\nonumber\\ 
              &&\hspace{0.68truecm} +\,\cv_{LR}\circ \check{G}_{\infty,R}\circ\cv_{RL}
                  \circ \check{G}_{\infty,L}\circ\check{T}_{LL}\nonumber\\
\check{T}_{RR}&=&\cv_{RL}\circ \check{G}_{\infty,L}\circ\cv_{LR}
\label{Tcomponents}\\
              &&\hspace{0.68truecm} +\,\cv_{RL}\circ \check{G}_{\infty,L}\circ\cv_{LR} 
                  \circ \check{G}_{\infty,R}\circ\check{T}_{RR}\nonumber\\
\check{T}_{LR}&=&\cv_{LR}
               +\cv_{LR}\circ \check{G}_{\infty,R}\circ\cv_{RL} \circ \check{G}_{\infty,L}\circ\check{T}_{LR}
\nonumber\\
\check{T}_{RL}&=&\cv_{RL}
               +\cv_{RL}\circ \check{G}_{\infty,L}\circ\cv_{LR} \circ \check{G}_{\infty,R}\circ\check{T}_{RL}.
\nonumber
\end{eqnarray}
The equations above depend only on the Green's functions $\check{G}_{\infty,L}$ and $\check{G}_{\infty,R}$ of the two 
uncoupled systems. Since the full Green's function $\check{G}_{ij}$ has been eliminated from the T-matrix equations
there are no Green's functions with spatial arguments in both systems. 
Together with the short range of $\cv_{LR,RL}$
this means that we can directly perform the quasiclassical $\xi$-integration on the T-matrix equations 
and substitute the quasiclassical Green's functions, $\cgia$,
for the full ones, $\check{G}_{\infty,i}$, above.

At the quasiclassical level, the Green's functions $\cgia(\pfhv;t,t^\prime)$ 
at the interface in equations (\ref{Tcomponents}) depend on
the position on the Fermi surface, $\pfhv$, and of two times $(t,t^\prime)$.
The coupling elements will be assumed to be time independent but may couple different points $\pfhv$ and
$\pfhv^\prime$ on the Fermi surfaces of the two conductors. The exact form of  
the $(\pfhv,\pfhv^\prime)$-dependence of
$\cv_{LR}$
is a degree of freedom in the model that allows us
to consider different types of transport through the interface.
We can now write down the equations for the quasiclassical $\ct$-matrix components 
\begin{eqnarray}
\ct_{LL}&=&\langle\cv_{LR}\ot \cgiR\ot\cv_{RL}\rangle_{\sfhv^{\prime\prime}}\nonumber\\ 
              &&\hspace{0.68truecm} +\,\langle\langle\cv_{LR}\ot \cgiR\ot\cv_{RL}
                  \ot \cgiL\ot\ct_{LL}\rangle_{\sfhv^{\prime\prime}}\rangle_{\sfhv^{\prime\prime\prime}}\nonumber\\
\ct_{RR}&=&\langle\cv_{RL}\ot \cgiL\ot\cv_{LR}\rangle_{\sfhv^{\prime\prime}}
\label{qcTequation}\\
              &&\hspace{0.68truecm} +\,\langle\langle\cv_{RL}\ot \cgiL\ot\cv_{LR} 
                  \ot \cgiR\ot\ct_{RR}\rangle_{\sfhv^{\prime\prime}}\rangle_{\sfhv^{\prime\prime\prime}}\nonumber\\
\ct_{LR}&=&\cv_{LR}
               +\langle\langle\cv_{LR}\ot \cgiR\ot\cv_{RL} \ot \cgiL\ot\ct_{LR}
\rangle_{\sfhv^{\prime\prime}}\rangle_{\sfhv^{\prime\prime\prime}}
\nonumber\\
\ct_{RL}&=&\cv_{RL}
               +\langle\langle\cv_{RL}\ot \cgiL\ot\cv_{LR} \ot \cgiR\ot\ct_{RL}
\rangle_{\sfhv^{\prime\prime}}\rangle_{\sfhv^{\prime\prime\prime}} ,
\nonumber
\end{eqnarray}
where we suppressed the explicit dependence of the functions on $\pfhv$ and on time variables. The
earlier $\circ$-product in Eq. (\ref{Tcomponents}) is replaced by the $\otimes$-product in the
quasiclassical expression. The $\otimes$-product
stands for an integration over a common time variable together with a normal matrix multiplication
in the combined Keldysh-Nambu and spin space\cite{Serene1983}. A left-over from the $\xi$-integration is the 
intermediate averaging over position on the Fermi surface as indicated by 
$\langle\cdots\rangle_{\sfhv}=\int_{FS} \cdots d\pfhv$. 
The quasiclassical $\ct$-matrix entering in to equation (\ref{transport}) 
is the forward scattering limit\cite{Thuneberg1984}
\[
\ct_{ij}=\ct_{ij}(\pfhv,\pfhv;t,t^\prime)
\]
of Eq. (\ref{qcTequation}) and in general it depends on two times $(t,t^\prime)$.

Following Ref. \onlinecite{Serene1983}, let us summarize the procedure for calculating the quasiclassical 
propagators in the presence of an interface:

\begin{enumerate}
\item To find $\check{g}_{\infty}$, we solve the conventional quasiclassical equations, the Eilenberger 
equation or the Usadel equation, for the uncoupled electrodes in equilibrium 
using hard-wall boundary conditions.

\item Use $\check{g}_{\infty}$ to solve the quasiclassical T-matrix equations (\ref{qcTequation}).

\item Solve the inhomogeneous quasiclassical equation (\ref{transport}) for the physical
 propagator $\check{g}$.

\item Use $\check{g}$ to calculate the ``smooth" self-energies $\ce_i$ and $\cD_i$
which enter the quasiclassical equations for $\check{g}_{\infty}$ and for $\check{g}$.
\end{enumerate}
The whole scheme amounts to a set of linear differential equations for $\check{g}_{\infty}$ and $\check{g}$, 
coupled in a non-linear way by the T-matrix and the self-energy equations. Substantial simplifications can be
achieved in the case of low transmissive tunnel barriers or point contacts. In these cases one can neglect the 
influence of the neighboring electrodes in the calculation of the self-energies and then the equations
for $\check{g}_{\infty}$ and $\ce_i$ and $\cD_i$ decouple.

\section{Josephson currents}
As a first application of the boundary condition
we calculate supercurrent through a variety of contacts connecting two superconducting reservoirs.
The current contribution from a given trajectory, $\pfhv$, and at a given energy, $\varepsilon$,
may be calculated directly
by integrating the transport equation (\ref{transport}) along the direction given by $\vfv(\pfhv)$.
This is easily seen as on the considered trajectory away from the contact
the physical propagator $\cg_i(\pfhv)$ coincides with the intermediate propagator $\cgia(\pfhv)$
calculated by the impenetrable surface condition. In absolute vicinity of the contact
only the source term in Eq. (\ref{transport}),
$\lbrack \ct_{ii}(\pfhv,\pfhv),\cgia(\pfhv) \rbrack \delta(\Rv-\Rv_{c})$, contributes and 
results in a jump in the Green's
function. The magnitude of this jump is given by integrating Eq. (\ref{transport}) 
over the interval $\rbrack 0_-,0_+\lbrack$. 
Performing the integral results in the scattered propagator

\begin{equation}
\cg_{i+}(\pfhv)=\cg_{i-}(\pfhv)-\frac{i}{\vf \cos \phi_i} \lbrack \ct_{ii}(\pfhv,\pfhv),\cg_{i-}(\pfhv) \rbrack ,
\label{greenjump}
\end{equation}

\noindent
where $\phi_i$ is the angle $\vfv(\pfhv)$ makes with the contact normal.
Note that $\cg_{i-}(\pfhv)\equiv\cgia(\pfhv)$ and to calculate $\cg_{i+}(\pfhv^\prime)$, i.e.
the propagator along the trajectory ($\pfhv^\prime$) coupled to ($\pfhv$) by pure surface scattering we must
solve for $\cg_{i+}(\pfhv^\prime)$ along the path given by $\vfv(\pfhv^\prime)$. The propagator at the contact,
computed in (\ref{greenjump}), may now be inserted in the current formula\lrule

\begin{equation}
j(T)=eN_F\int \frac{d\varepsilon}{4\pi i}\; {\rm Tr} \langle \vfv(\pfhv) g^K(\pfhv;\varepsilon) \rangle_{\pfhv}
    =eN_F\int \frac{d\varepsilon}{4\pi i}\;  \langle j^K_\varepsilon(\pfhv) \rangle_{\pfhv} ,
\label{jocurrent}
\end{equation}

\noindent
with $N_F$ the density of states at  the Fermi level in the normal state.
Since the Josephson current is an equilibrium property, the Keldysh Green's function components of $\cg$ 
are in this case simply related to the Retarded (R) and Advanced (A) ones as 
$\hat g^K=(\hat g^R-\hat g^A) \tanh (\varepsilon/2 T)$, and with
$(\hat g^A(\pfhv;\varepsilon))^{\dagger}=\hat\tau_3\hat g^R(\pfhv;\varepsilon)\hat\tau_3$,
we get the energy and trajectory resolved
current contribution across the contact, evaluated in the left superconductor, as
\begin{equation}
j^K_\varepsilon(\pfhv)= {\rm Im} \bigg\lbrack {\rm Tr}
 \bigg\lbrace\; i\hat{\tau}_3\; \lbrack \hat{t}_{LL}^R,\hat{g}_{\infty,L}^R \rbrack \bigg\rbrace
                        \bigg\rbrack\,\tanh(\frac{\varepsilon}{2 T}).
\label{currentcontrib}
\end{equation}
The lonely first intermediate Green's function, $\cgia(\pfhv)$, in Eq. (\ref{greenjump}) explicitly drop
out of the current in the angle average since they obey the impenetrable surface boundary condition.

So far no reference have been made to the modeling of the contact and
the $\hat t$-matrix element $\hat t^R_{LL}$. The contact model depends on the choice of the momentum
dependence of the coupling elements, $v^{\dagger}_{RL}=v_{LR}=v(\pfhv,\ppfhv)$. Two extreme models for
the $(\pfhv,\ppfhv)$-dependence will be considered: a totally disordered contact,
$v(\pfhv,\ppfhv)=v$, i.e. the coupling across the contact retains no memory of the momentum direction, 
and a momentum conserving contact with $v(\pfhv,\ppfhv)=v \delta(\pfhv-\ppfhv)$. The $\hat t$-matrix equations 
for the two types of contact, dropping superfluous indexing, read
\begin{equation}
\hat t= \hat{v}\, \langle\hat g_R(\pfhv)\rangle_{\sfhv}\, \hat{v}^{\dagger}
      + \hat{v}\, \langle\hat g_R(\pfhv)\rangle_{\sfhv}\, \hat{v}^{\dagger}\,
            \langle\hat g_L(\pfhv)\rangle_{\sfhv}\, \hat t
\label{dispc}
\end{equation}
for the disordered contact and
\begin{equation}
\hat t(\pfhv)= \hat{v}\, \hat g_R(\pfhv) \hat{v}^{\dagger}
             + \hat{v}\, \hat g_R(\pfhv)\, \hat{v}^{\dagger}\, \hat g_L(\pfhv)\, \hat t(\pfhv)
\label{conpc}
\end{equation}
for the momentum conserving contact. For either model, the $\hat t$-matrix equation above is simple to 
invert after inserting the retarded Green's functions $\hat g_R(\pfhv)$ and $\hat g_L(\pfhv)$
\begin{equation}
\hat g(\pfhv)_{R(L)}=\left(\begin{array}{cc}g_{R(L)} &f_{R(L)}e^{\pm i\chi/2}\\
-\tilde f_{R(L)}{\rm e}^{\mp i\chi/2}&-g_{R(L)}\end{array}\right) ,
\label{gchi}
\end{equation}
with the phase difference $\chi$ across the junction and the upper (lower) signs of the phase refer to 
the right (left) electrode. In equilibrium the T-matrix equation is simply an algebraic equation in energy
space, which can be trivially inverted. The energy and trajectory resolved current is written
\begin{equation}
j_\varepsilon(\pfhv)= {\rm Im} \bigg\lbrack{{ i\, {\cal D}\; ( \fr \tl e^{i\chi}-\fl \tr e^{-i\chi}) }\over
{2-{\cal D}-{\cal D}\gr\gl +\frac{{\cal D}}{2}(\fr \tl e^{i\chi}+\fl \tr e^{-i\chi})}}\bigg\rbrack
                        \tanh(\frac{\varepsilon}{2 T}).
\label{josephcurrent}
\end{equation}
\rrule
Here we have traded in the coupling strength, $v$, for the 
transmission coefficient ${\cal D}$. The two are simply related as \cite{MartinRodero1994}
\begin{equation}
{\cal D}={{4 |v|^2}\over{(1+|v|^2)^2}}.
\label{transmissioncoeff}
\end{equation}

\noindent
The scalar coupling element, $v$, is from now on a dimensionless quantity. We have normalized the original
coupling element in units of $(1/\pi N_F)$ to get rid of the different prefactors that appear in the
angle averages in the T-matrix equation (\ref{qcTequation}).

%=================================================================================================
\subsection{Josephson current between two s-wave superconductors}
Assuming that the two electrodes both are s-wave superconductors, 
we have the Green's functions on either side of the contact
\begin{equation}
\hat g(\pfhv)_{R(L)}=-\frac{\pi}{\Omega}\left(\begin{array}{cc}\varepsilon &\Delta {\rm e}^{\pm i\chi/2}\\
-\Delta {\rm e}^{\mp i\chi/2}&-\varepsilon\end{array}\right) ,
\label{swaveg}
\end{equation}
where $\Omega=\lbrack\Delta^2-\varepsilon^2\rbrack^{\frac{1}{2}}$. 
Since the s-wave superconductor is isotropic it does
not matter which model for the coupling, (\ref{dispc}) or (\ref{conpc}), we choose. Additionally, in the absence 
of surface depairing effects it is
sufficient to know the bulk Green's functions (\ref{swaveg})
to calculate the current contribution (\ref{currentcontrib}).
Using Eq. (\ref{josephcurrent}) we find the known result that the Josephson current is carried by 
junction states\cite{Furusaki1990,Beenakker1991} located at 
$\varepsilon_J(\chi,T)=\pm\Delta(T) \lbrack 1-{\cal D} \sin^2(\chi/2)\rbrack^{\frac{1}{2}}$. 
The total current is the sum of all contributions and reads
\begin{equation}
j(T)=eN_F{\cal D}\,{{\pi \Delta(T) \sin\chi}\over{\lbrack 1-{\cal D} \sin^2\frac{\chi}{2}\rbrack^{\frac{1}{2}}}} 
\tanh({\varepsilon_J(\chi,T)\over{2T}}).
\label{sjoseph}
\end{equation}
In equation (\ref{sjoseph}) it should be noted that a second temperature dependence enters
via the the temperature dependent gap, $\Delta(T)$.

%=================================================================================================
\subsection{Josephson current between two d-wave superconductors}
To emphasize the modeling of the $(\pfhv,\ppfhv)$-dependence of the coupling across the junction
and the importance of
using the correct surface Green's functions,
we now study the current-phase relation of two d-wave superconductors. A realization of a d-wave
order parameter is $\Delta_{\sfhv}=\Delta \cos 2(\phi_{\sfhv}-\alpha)$.
The magnitude and sign of $\Delta_{\sfhv}$ depends on the position on the Fermi circle and this is measured 
by the angle, $\phi_{\sfhv}$, the angle $\pfhv$ makes with
the crystal $\hat a$-axis. The angle $\alpha$ tracks the relative junction-to-crystal $\hat a$-axis orientation.
If $\alpha=\pm\pi/4$ and specular quasiparticle 
scattering at the interface
is assumed the order parameter seen along a trajectory changes sign at the surface and 
an Andreev bound state forms at zero energy for every trajectory $\pfhv$ [\onlinecite{Hu1994}].
We will stick with the junction realization $\alpha=\pm\pi/4$.
To incorporate the effect of these surface states into the current-phase relation
one must use the surface Green's functions\cite{Barash1996,Fogelstrom1997}
\begin{equation}
\hat g(\pfhv)_{R(L)}=\frac{\pi}{\varepsilon}
\left(\begin{array}{cc}\Omega_{\sfhv} &i s_R \Delta_{\sfhv} {\rm e}^{\pm i\chi/2}\\
i s_R \Delta_{\sfhv}  {\rm e}^{\mp i\chi/2}&-\Omega_{\sfhv} \end{array}\right) ,
\label{dwaveg}
\end{equation}
where $\Omega_{\sfhv}=\lbrack\Delta_{\sfhv}^2-\varepsilon^2\rbrack^{\frac{1}{2}}$. The factor
$s_R$ discriminates between two types of junction\cite{Barash1996,Barash2000}:
$s_R=-1\; (\alpha_R=\alpha_L$) is referred to as a symmetric, and
$s_R=1\; (\alpha_R=-\alpha_L$) as a mirror junction. 
The convention for signs of the phase, $\chi$,
are as for the s-wave superconductor. It should be said before proceeding that we are 
neglecting the pair-breaking effect of the surface\cite{Buchholtz1995} and we assume constant order parameters
up to the interface. This is for the sake of simple illustration and for the comparison of 
analytical results with other boundary conditions.

Turning to the $\hat t$-matrix equations and starting with the diffusive model of the point contact,
one immediately find that the Josephson current is zero. This follows from the vanishing average,
$\langle \Delta_{\sfhv}\rangle_{\sfhv}=0$. Due to this property the anomalous propagators vanish
and therefore the Josephson current as well (see Eq. (\ref{josephcurrent})).
In the opposite limit, using the momentum-conserving model (\ref{conpc}), the Josephson current is not zero.
After inverting the $\hat t$-matrix equation and evaluating the commutator in (\ref{currentcontrib}),
we find the energy resolved current at $\pfhv$ 
\begin{equation}
j_{\varepsilon}(\pfhv)=\pm{\rm Im} \bigg\lbrack {\cal D} \Delta^2_{\sfhv}
\bigg( {{\sin \chi}\over{\varepsilon^2-\varepsilon^2_{J}(\chi;\pfhv)}}\bigg)\,
\bigg\rbrack \tanh(\frac{\varepsilon}{2 T}) .
\end{equation}
The sign of $j_{\varepsilon}(\pfhv)$ is (+) for a mirror and (-) for a symmetric junction.
As in the case of the s-wave junction we have junction states carrying the Josephson current. 
The position of these states depend on the type
of junction in the following way
\begin{equation}
\varepsilon_{J}(\chi;\pfhv)=\pm\sqrt{{\cal D}}\, |\Delta_{\sfhv}| \times \bigg\lbrace\begin{array}{l}
|\sin\frac{\chi}{2}|,\quad\mbox{mirror}\\*[0.2truecm]
|\cos\frac{\chi}{2}|,\quad\mbox{symmetric} .
\end{array}
\end{equation}
These junction states were first found by Riedel and Bagwell\cite{Riedel1998} using a scattering approach
and, independently, by Barash\cite{Barash2000} using what is known as the Zaitsev boundary
condition\cite{Zaitsev1984}.
Performing the integral over the energy we write the trajectory resolved current-phase relations
\begin{equation}
j(\pfhv)=\pm 2 \pi \sqrt{{\cal D}}\, |\!\Delta_{\sfhv}\!|\! \times\! \Bigg\lbrace\begin{array}{l}
\cos \frac{\chi}{2}\tanh({{\sqrt{{\cal D}}\, |\!\Delta_{\sfhv}\!| \sin\frac{\chi}{2}}\over{2 T}})\\*[0.3truecm]
\sin \frac{\chi}{2}\tanh({{\sqrt{{\cal D}}\, |\!\Delta_{\sfhv}\!| \cos\frac{\chi}{2}}\over{2 T}})
\end{array}
\label{tresdjoseph}
\end{equation}
for the two junction types, the upper being the mirror and the lower the symmetric one. The total current is the
trajectory average of $j(\pfhv)$ multiplied by $eN_F$.

The main purpose of this section was to show that the quasiclassical version of the point contact
coupling of two electrodes is simple to use and can recover results known in literature. In case of
unconventional superconducting electrodes results depend in a crucial way on how the contact is
modeled. This gives the Hamiltonian boundary condition an advantage in flexibility
to the conventional Zaitsev boundary condition which coincides with the momentum conserving contact 
(\ref{conpc}) introduced above.
%=================================================================================================
\subsection{Josephson current through a spin active interface}

As another illustration of the flexibility of this method we
extend the discussion to currents through spin active interfaces, i.e. interfaces which flip
the spin of the incident electrons either by spin-dependent scattering within the interface, or
by a difference in spin-orbit coupling on either side of the interface.
The general boundary conditions that connect the quasiclassical propagators for superconducting metals
across magnetically active interfaces were introduced by Millis, Rainer and Sauls \cite{Millis1988}. Recently one of
the authors \cite{Fogelstrom2000} derived the explicit solution of these boundary conditions for 
equilibrium Green functions.
In order to compare with this solution, as in Ref. [\onlinecite{Fogelstrom2000}],
we shall analyze in this section the Josephson
current through a contact of two isotropic s-wave superconductors connected 
through a small magnetically active junction.

In order to accommodate the spin dependence we enlarge our space in such a way that every quantity is now
a $2 \times 2$ matrix in spin space. In particular, the coupling elements are spin dependent and 
adopt the following general form
\begin{equation}
\hat{v}_{LR} = \hat{v}^{\dagger}_{RL}  = \left(\begin{array}{cc} v & 0 \\
 0 & v^{\dagger} \end{array}\right) \;\mbox{where}\;
 v = \left(\begin{array}{cc}  v_{\uparrow,\uparrow} & v_{\uparrow,\downarrow} \\
v_{\downarrow,\uparrow}  & v_{\downarrow,\downarrow} 
\end{array}\right) .
\end{equation}
Let us stick to the case of S/F/S junction analyzed in Ref. [\onlinecite{Fogelstrom2000}]. 
In this case $F$ stands for a small 
ferromagnetic particle or grain. This ferromagnetic material is treated as a partially transparent barrier which 
transmits the two spin projections differently. For spin-active interfaces the different components of the 
$S$-matrix, $S_{ij}$ are $2\times2$ spin matrices. To proceed further a specific $S$-matrix was chosen 
in Ref. [\onlinecite{Fogelstrom2000}] to model 
the magnetic barrier
\begin{equation}
 \hat S=\left(\begin{array}{cr} S_{11} & S_{12} \\ S_{21}& S_{22}\end{array}\right)=
      \left(\begin{array}{cr} r & t \\ t& -r\end{array}\right)\exp( i \Theta \sigma_3) ,
\label{S-matrix}
\end{equation}
where $\sigma_j$ notes the Pauli-matrices spanning spin space and parameters $(t,r)$ are the usual transmission 
and reflection coefficients. The S-matrix (\ref{S-matrix}) is one of the simplest choices that allows a
variable degree of spin mixing at the interface and the spin mixing is parameterized by the spin-mixing angle 
$\Theta$.  By this construction $\hat S$ only violates spin conservation, i.e. it does not commute with the 
quasiparticle spin operator $\sigma$. The angle $\Theta$ will be considered as a phenomenological parameter 
independent of the trajectory direction (for more details see Tokuyasu {\it et al} Ref. [\onlinecite{TSR1988}]).
Within the approach presented of this paper, one can easily model the previous S-matrix with
a spin dependent coupling
\begin{equation}
\hat{v}_{LR} = \hat{v}^{\dagger}_{RL}  = v \left(\begin{array}{cc}  \exp( i \Theta \sigma_3) & 
0 \\ 0 & \exp( -i \Theta \sigma_3) 
\end{array}\right).
\end{equation}
\lrule

Again the quasiclassical surface Green functions are the inputs in this approach, i.e. the Green's functions 
at the interface calculated with impenetrable wall boundary conditions. The spin-active boundary condition must
be used at the contact also for the reflecting surface\cite{Fogelstrom2000}.
In this simple case of a magnetic barrier
the resulting spin-dependent propagators can be written in a $4\times 4$ block-diagonal form
\begin{equation}
\hat g_{block}(\pfhv;\Theta)=
\left(\begin{array}{cc} 
\hat g(\pfhv;\Theta) & 0 \\
0   & \hat g(\pfhv; -\Theta)\end{array}\right)=
\left(\begin{array}{cccc} g_{\uparrow\uparrow}& f_{\uparrow\downarrow}&0&0\\
                          \tilde f_{\downarrow\uparrow}&\tilde g_{\downarrow\downarrow}&0&0\\
                         0&0&g_{\downarrow\downarrow}&-f_{\downarrow\uparrow}\\
                         0&0&-\tilde f_{\uparrow\downarrow}&\tilde g_{\uparrow\uparrow}\end{array}\right) ,
\end{equation}
\noindent
where the electron and anomalous parts of $\hat g(\pfhv;\Theta)$ in the upper left corner
of $\hat g_{block}(\pfhv;\Theta)$ can be written as
\begin{eqnarray}
g_{\uparrow\uparrow}(\Theta) &=& - \pi \frac{\varepsilon^R \cos(\Theta/2)-\Omega \sin(\Theta/2)}
                                             {\varepsilon^R \sin(\Theta/2)+\Omega \cos(\Theta/2)}\\
\nonumber
f_{\uparrow\downarrow}(\Theta) &=&\;\; \pi \frac{\Delta {\rm e}^{-i(\Theta\mp\chi)/2}}
                                          {\varepsilon^R \sin(\Theta/2)+\Omega \cos(\Theta/2)}\\
\tilde f_{\downarrow\uparrow}(\Theta) &=&-\pi  \frac{\Delta {\rm e}^{ i(\Theta\mp\chi)/2}}
                                            {\varepsilon^R \sin(\Theta/2)+\Omega \cos(\Theta/2)}
\nonumber
\end{eqnarray}
for trajectories with $(\pfhv\cdot\nhv>0)$.  
For trajectories with reversed momentum, i.e. $(\pfhv\cdot\nhv<0)$,
the phase factor $\exp\lbrack\mp i(\Theta\mp\chi)/2\rbrack$ for functions $f_{\uparrow\downarrow}(\Theta)$
and $\tilde f_{\downarrow\uparrow}(\Theta)$ goes to $\exp\lbrack \pm i(\Theta\pm\chi)/2\rbrack$.
Above, as in the earlier examples, the phase difference, 
$\chi$, between the two reservoirs is included 
and the upper (lower) signs refer two the right (left) reservoir. The components
of the propagator in the lower right corner of $\hat g_{block}$ are simply related to those in the upper left
corner by the replacement $\Theta \rightarrow -\Theta$.

The angle $\Theta$ induces a mixing of the two otherwise separated spin bands. 
It is easy to see that the resulting density of states at the 
interface has Andreev bound states inside the gap. 
These states are located 
at $\varepsilon_{b,\uparrow (\downarrow)} = \pm \Delta \cos(\Theta/2)$, with $+(-)$ for the spin-up (-down) branch.
The existence of the sub-gap states alters the Josephson current-phase relation radically. The contribution
to the current from the energy $\varepsilon$, the trajectory $\pfhv$, and spin band $\uparrow (\downarrow)$ 
may be calculated directly from expression (\ref{josephcurrent}). Keeping in mind that the
phase that enters in Eq. (\ref{josephcurrent}) is the phase difference over the contact we write
\begin{equation}
j_{\varepsilon,\uparrow}(\pfhv;\Theta,\chi) =  {\rm Im} \bigg\lbrack {\cal D} \Delta^2
\bigg( {{\sin \chi}\over{\left[ \Omega \cos(\Theta/2) + \varepsilon \sin(\Theta/2) \right]^2
- {\cal D} \Delta^2 \sin^2(\chi/2) }}\bigg)\,
\bigg\rbrack \tanh(\frac{\varepsilon}{2 T}) 
\label{spectral-current}
\end{equation}
for the current carried by the spin-up band. The current carried by the spin-down band is given
simply as $j_{\varepsilon,\downarrow}(\pfhv;\Theta,\chi)= j_{\varepsilon,\uparrow}(\pfhv;-\Theta,\chi)$.
At a finite superconducting phase difference, the original two interface Andreev bound states are 
split up into four current carrying states located inside the gap at positions  
\begin{equation}
\varepsilon_J = \pm \Delta \left[ \cos^2(\Theta/2) - {\cal D} \cos(\Theta) \sin^2(\chi/2) \pm
\sqrt{{\cal D}} \sin(\Theta) \sin(\chi/2) \sqrt{1 - {\cal D} \sin^2(\chi/2)} \right]^{1/2}.
\end{equation}
\rrule
These states give the total contribution to the current and their positions change from the tunnel
regime, $\varepsilon_J = \pm \Delta \cos(\Theta/2)$, to $\varepsilon_J = \pm \Delta \cos[(\chi \pm \Theta)/2)]$
at perfect transmission.
\begin{figure}
\centerline{\rotate[r]{\epsfxsize=0.37\textwidth{\epsfbox{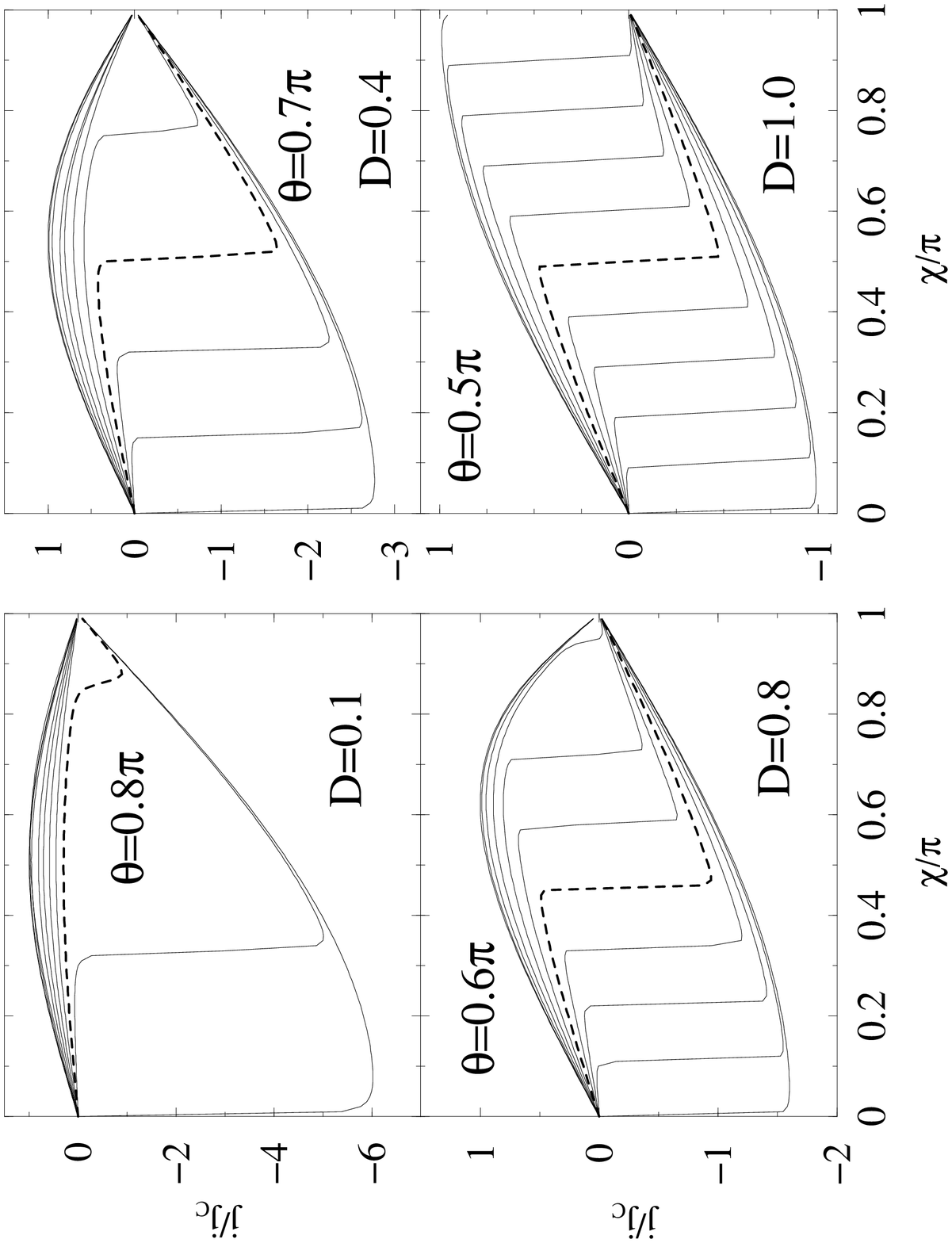}}}}
\caption{Zero-temperature supercurrent-phase relation for the S/F/S contact considered in this section.
The four different panels correspond to ${\cal D}=0.1, 0.4, 0.8$, and $1.0$. The spin-mixing angle $\Theta$ is 
varied from top to bottom from $0$ to $\pi$ in steps of $\frac{\pi}{10}$. The dashed lines indicate the value
of $\Theta$ for which the contacts become $\pi$ junctions. The supercurrent is normalized in units of the 
critical current density, $j_C$, for the corresponding transmission and zero spin-mixing angle.}
\label{SpinCP}
\end{figure}
In figure \ref{SpinCP} we show the current-phase relation for a set of transparencies ${\cal D}=0.1, 0.4, 0.8$,
and $1.0$. In each panel
the spin-mixing angle $\Theta$ is varied from $0$ to $\pi$ in steps of $\frac{\pi}{10}$. As seen, for each 
value of ${\cal D}$ there is a range $\Theta > \Theta_c$ where the junctions are $\pi$-junctions. 
For small ${\cal D}$, the critical spin-mixing angle $\Theta_c$ is close to $\pi$ and with increasing ${\cal D}$,
$\Theta_c$ increases towards $\frac{\pi}{2}$.  The magnitude of the critical current is for all but the perfect
transmission junction very asymmetric for $0$ and $\pi$ junctions. In figure \ref{Jstates} we show the energy 
resolved spectral current (\ref{spectral-current}) for ${\cal D}=0.4$ and $\Theta=0.7\pi$ at different phase
differences, $\chi$, over the junction. At small phase differences the junction
state initially at $\varepsilon_J(\chi=0) = \pm \Delta \cos(\Theta/2) $ splits in to two states
carrying current in opposite directions. This gives a small but positive current as seen in the 
corresponding current-phase
relation in figure \ref{SpinCP}. As the phase difference is increased the two states dispersing with phase towards
$\varepsilon=0$ from either side eventually cross at $\varepsilon=0$. Above this phase difference, 
both current-carrying states at $\varepsilon<0$ ($\varepsilon>0$)
give current in the same direction and the magnitude of the current increases abruptly.

To conclude this section, let us stress that these results reproduce the result obtained in Ref. 
\onlinecite{Fogelstrom2000}, showing again the versatility of the boundary conditions introduced in this work.
\begin{figure}
\centerline{\rotate[r]{\epsfxsize=0.4\textwidth{\epsfbox{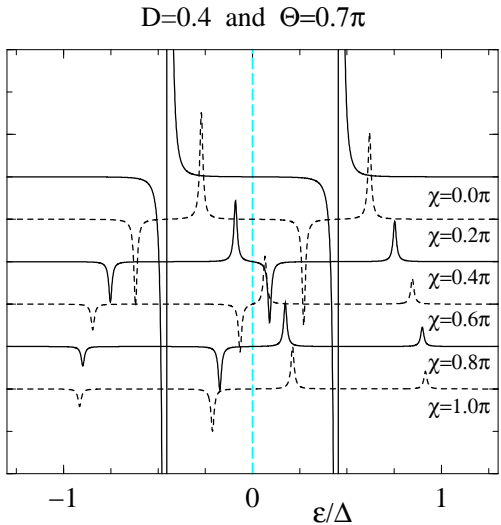}}}}
\caption{Energy resolved spectral current for the S/F/S contact with ${\cal D}=0.4$ and $\Theta=0.7\pi$ (see 
Fig. \ref{SpinCP} right upper panel). This figure shows the total spectral current, sum of both spin contributions,
as a function of energy according to Eq. (\ref{spectral-current}), without the thermal factor. The different
curves show the evolution of the four current-carrying Andreev bound states inside the gap with the superconducting
phase difference. The curves are shifted vertically for clarity.}
\label{Jstates}
\end{figure}

\section{Solving the T-matrix equation at an applied voltage}
In this section we shall describe how the T-matrix equation (\ref{qcTequation}) can be solved in the case of
a voltage-biased superconducting contact. As a constant bias $V$ is applied across a junction 
between two superconductors, the phase difference oscillates with time according to the Josephson
relation: $\phi(t) = \phi_0 + \omega_0 t$, where $\omega_0 = 2eV/\hbar$ is the Josephson frequency. 
This makes that every Green's function and T-matrix component depend on two time arguments. We show in this 
section how the time convolutions in the T-matrix equation can be handled.

As we shall show in next section the current can be expressed, for instance, only in terms of the 
advanced and retarded components of $\ct_{LR}$. Thus, we concentrate ourselves in these components
whose equations can be written as (see Eq. 9)

\begin{eqnarray}
\hat{t}^{R,A}_{LR}(t,t^{\prime}) & = &  \hat{v}_{LR} + \int dt_1 \int dt_2 \; \hat{v}_{LR} \times 
\nonumber \\
& & \hat{g}^{R,A}_{R}(t,t_1) \; \hat{v}_{RL} \; \hat{g}_{L}^{R,A}(t_1,t_2) \; 
\hat{t}^{R,A}_{LR}(t_2,t^{\prime}).
\label{tmatrixLR}
\end{eqnarray}

\noindent
Here, we have written explicitly the time convolutions for the sake of clarity and we have 
omitted the $\pfhv$ integrations, since they do not affect the time convolutions. In this
expression every quantity is a $4 \times 4$ matrix in Nambu and spin space, and from now to the end 
of this section we get rid of the superindexes $R,A$, since the equations for these two components are
formally identical. We also drop the $\infty$ subindex, since all propagators in (\ref{tmatrixLR})
are propagators of the separated electrodes. 
The electrode Green's functions entering Eq. (\ref{tmatrixLR})
take the form: $\hat{g}_{j}(t,t^\prime) = \hat{U}^{\dagger}_j(t) \hat{g}_{j}(t-t^\prime) 
\hat{U}_j(t^\prime)$, where $j=R,L$ and $\hat{U}_j(t) = \mbox{exp} \left[i \phi_i(t) \hat{\tau_3} /2 
\right]$, $\phi_j(t)$ being the phase of the $j$th superconductor. In this expression, $\hat{g}_{j}(t) 
= \int \hat{g}_{j}(\epsilon) \mbox{exp}(-i\epsilon t) \; d\epsilon/2\pi$.

We use the transformation generated by $\hat{U}_j(t)$ to transfer the time dependence from
the Green's functions to the hopping elements 

\begin{eqnarray}
\hat{t}_{LR}(t,t^{\prime}) & = & \hat{v}_{LR}(t) \delta(t-t^\prime) + \int dt_1 \int dt_2 \;
\hat{v}_{LR}(t) \nonumber \\
& & \hat{g}_{R}(t-t_1) \; \hat{v}_{RL}(t_1) \; \hat{g}_{L}(t_1-t_2) \; \hat{t}_{LR}(t_2,t^{\prime}),
\label{time-Tmatrix}
\end{eqnarray}

\noindent
where $\hat{v}_{LR}(t) = \hat{U}_L(t) \hat{v}_{LR} \hat{U}^{\dagger}_R(t) = \hat{v}_{RL}^{\dagger}(t) =
v \;\mbox{exp} \left[i \phi(t) \hat{\tau}_3 /2 \right]$. One can easily show that all physical 
properties of the system are invariant under this gauge transformation. Thus, we shall usually consider 
the T-matrix equation in this gauge, i.e. in which the hopping elements are time-dependent
and the electrode Green's functions only depend on the time difference.

In order to solve Eq. (\ref{time-Tmatrix}) it is more convenient to work in energy space where it becomes an 
algebraic equation. Thus, we Fourier transform the T-matrix with respect to the temporal arguments

\begin{equation}
\hat{t}_{LR}(t,t^{\prime}) = \frac{1}{2 \pi} \int d\epsilon \; \int
d\epsilon^{\prime} \; e^{-i \epsilon t} e^{i \epsilon^{\prime} t^{\prime}}
\; \hat{t}_{LR}(\epsilon,\epsilon^{\prime}).
\end{equation}

\noindent
It is easy to convince oneself that, due to the
special time dependence of the coupling elements, the T-matrix admits a Fourier
expansion of the form

\begin{equation}
\hat{t}_{LR}(t,t^{\prime}) = \sum_{n} e^{i n \phi(t^{\prime})/2} \int \frac{d\omega}{2 \pi}
\; e^{-i \epsilon (t-t^{\prime})} \; \hat{t}_{LR}(\epsilon, \epsilon+n eV) .
\label{Fourier-expansion}
\end{equation}

\noindent
In other words, Fourier transforming Eq. (\ref{time-Tmatrix}) one can show that
$\hat{t}_{LR}(\epsilon,\epsilon^{\prime})$ satisfies the following relation

\begin{equation}
\hat{t}_{LR}(\epsilon,\epsilon^{\prime}) = \sum_{n}
\hat{t}_{LR}(\epsilon,\epsilon + n eV) \;
\delta(\epsilon -\epsilon^{\prime} + n eV) .
\end{equation}

The problem of the calculation of the
current can be reduced to the evaluation of the Fourier components
$\hat{t}_{nm}(\epsilon) \equiv \hat{t}_{LR}(\epsilon+neV, \epsilon + m
eV)$. As it can be seen Fourier transforming Eq. (\ref{time-Tmatrix}), these
components fulfill the following (infinite) set of algebraic linear equations

\begin{equation}
\hat{t}_{nm} = \hat{v}_{nm} \delta_{n,m \pm 1} + \hat{{\cal{E}}}_{n} \hat{t}_{nm} + 
\hat{{\cal{V}}}_{n,n-2} \hat{t}_{n-2,m} + \hat{{\cal{V}}}_{n,n+2} \hat{t}_{n+2,m} ,
\label{Linear-system}
\end{equation}

\noindent
where $\hat{v}_{m-1,m} = v (\hat{1} + \hat{\tau}_3)/2$, $\hat{v}_{m+1,m} = v^{\dagger} 
(\hat{1} - \hat{\tau}_3)/2$, and the matrix coefficients $\hat{{\cal{E}}}_{n}$ and $\hat{{\cal{V}}}_{n,m}$ can
be expressed in terms of the Green's functions of the uncoupled electrodes, as

\begin{eqnarray}
\hat{{\cal{E}}}_{n} & = & \left(
\begin{array}{cc}
v \; g_{R,n+1} \; v^{\dagger} \; g_{L,n}  &   
v \; g_{R,n+1} \; v^{\dagger} \; f_{L,n}   \\
v^{\dagger} \; g_{R,n-1} \; v \; \tilde{f}_{L,n}  & 
v^{\dagger} \; g_{R,n-1} \; v \; g_{L,n}
\end{array} \right) \nonumber \\
\hat{{\cal{V}}}_{n,n+2} & = & -v f_{R,n+1} v \left(
\begin{array}{cc}
 \tilde{f}_{L,n+2}  &   g_{L,n+2}      \\
 0    &   0
\end{array} \right) \nonumber \\
\hat{{\cal{V}}}_{n,n-2} & = & -v^{\dagger} \tilde{f}_{R,n-1} v^{\dagger} \left(
\begin{array}{cc}
    0      &   0      \\
g_{L,n-2}   &  f_{L,n-2}
\end{array} \right)  ,
\end{eqnarray}

\noindent
In these equations the short-hand notation $g_{i,n}=g_i(\epsilon+ neV)$ for the $2 \times 2$ 
spin-dependent propagators has been used. Notice that the set of linear equations (\ref{Linear-system})
are analogous to those describing a tight-binding chain with nearest-neighbor
hopping parameters $\hat{{\cal{V}}}_{n,n+2}$ and $\hat{{\cal{V}}}_{n,n-2}$. A solution can then
be obtained by standard recursive techniques (see Ref. [\onlinecite{Cuevas1996}] for details).

Finally, the $\pfhv$-dependence of the Green's function in Eq. (\ref{Linear-system}) depends on our choice of
the contact model. Thus for instance, for the disordered case the Green's functions appearing
in Eq. (\ref{Linear-system}) are the angle averaged ones, while for the case of a momentum conserving contact
we must include the trajectory dependent Green's functions (see Eqs. 13-14).

\subsection{Current at finite voltage}

As commented in a previous section, the current contribution from a given trajectory may be calculated 
directly by integrating the transport equation (\ref{transport}) along the trajectory over the discontinuity
given by the source term. Thus, the time-dependent current reads as

\begin{equation}
j(t)=eN_F \langle j(\pfhv,t) \rangle_{\pfhv},
\end{equation}
\noindent
where the contribution of given trajectory with momentum $\pfhv$ can be written as

\begin{equation}
j(\pfhv,t)= {\rm Tr} \left\{ \hat{\tau}_3\; 
\lbrack \check{t}_{LL},\check{g}_{\infty,L} \rbrack^K_{\otimes} \right\} .
\end{equation}

This expression can be greatly simplified as follows. First, the Keldysh components of 
the T-matrix can be eliminated in favor of the advanced and retarded components using the relation
$\tilde{t}^K = \tilde{t}^R \otimes \tilde{g}^K_{\infty} \otimes \tilde{t}^A$. On the other hand, the four elements
of the enlarged space are not independent. For instance, it is easy to show the following relations:
$\check{t}_{LR} = \left(1 + \check{t}_{LL} \otimes \check{g}_{\infty,L} \right) \otimes \check{v}_{LR}$ and
$\check{t}_{RL} = \check{v}_{RL} \otimes \left(1 + \check{g}_{\infty,L} \otimes \check{t}_{LL} \right)$.
Using these relations it is rather straightforward to show that the current can be written as \cite{Cuevas1996}
\lrule

\begin{eqnarray}
j(\pfhv,t) & = &  \mbox{Tr} \left[ \hat{\tau}_3
\left( \hat{t}_{LR}^R \otimes \hat{g}_{R}^{K} \otimes \hat{t}_{RL}^{A} \otimes \hat{g}_{L}^A
- \hat{g}_{L}^{R} \otimes \hat{t}_{LR}^R \otimes \hat{g}_{R}^{K} \otimes \hat{t}_{RL}^{A}
\nonumber \right. \right. \\ && \hspace{0.9cm} \left. \left. +
\hat{g}_{R}^{R} \otimes \hat{t}_{RL}^{R} \otimes \hat{g}_{L}^{K} \otimes \hat{t}_{LR}^{A}
- \hat{t}_{RL}^{R} \otimes \hat{g}_{L}^{K} \otimes \hat{t}_{LR}^{A} \otimes \hat{g}_{R}^{A}
\right) \right] ,
\end{eqnarray}
\noindent
where we have dropped the symbol $\infty$, since from now on in this section the only Green functions
which will appear are the surface Green functions.

Taking into account the Fourier expansion of the T-matrix (see Eq. \ref{Fourier-expansion}), the
current adopts the form

\begin{equation}
j(\pfhv,t) = \sum_{m=-\infty}^{\infty} j_m(\pfhv) e^{im \phi(t)} ,
\end{equation}

\noindent
where the different Fourier current components can be expressed in terms of the Fourier components 
of the harmonics $\hat{t}_{nm}(\epsilon) \equiv \hat{t}(\epsilon+neV,\epsilon+meV)$ as follows

\begin{eqnarray}
j_m(\pfhv) & = &  \int d\epsilon \sum_n \mbox{Tr} \left[
\hat{\tau}_3 \left( \hat{t}_{LR,0n}^R \hat{g}_{R,n}^K \hat{t}_{RL,nm}^A \hat{g}_{L,m}^A - 
\hat{g}_{L,0}^R \hat{t}_{LR,0n}^R \hat{g}_{R,n}^K \hat{t}_{RL,nm}^A
\right. \right. \nonumber \\
& & \hspace{2.3cm} \left. \left. + \hat{g}_{R,0}^R \hat{t}_{RL,0n}^R \hat{g}_{L,n}^K \hat{t}_{LR,nm}^A 
- \hat{t}_{RL,0n}^R \hat{g}_{L,n}^K \hat{t}_{LR,nm}^A \hat{g}_{R,m}^A \right) \right].
\end{eqnarray}

This is the general expression of the ac Josephson effect in a superconducting
contact. This indicates that in the case of constant bias voltage there appear
alternating components which oscillate not only with the Josephson frequency
$\omega_0$, as in the case of a tunnel junction, but also with all its
harmonics. When the voltage tends to zero all the harmonics sum up to yield the
supercurrent.

We can further simplify the expression of the current harmonics, $j_m$. Using
the equations of the T-matrix components it can be shown that the general relation 
$\hat{t}^{A,R}_{RL,nm}(\epsilon) = \hat{t}^{R,A\dagger}_{LR,mn} (\epsilon)$ holds. Thus, we can simply express 
the current in terms of harmonics $\hat{t}^{R,A}_{LR,nm} \equiv \hat{t}^{R,A}_{nm}$ as follows

\begin{eqnarray}
j_m(\pfhv) & = & \int d\epsilon \sum_n \mbox{Tr} \left[
\hat{\tau}_3 \left( \hat{t}_{0n}^R \hat{g}_{R,n}^K \hat{t}_{mn}^{R\dagger} \hat{g}_{L,m}^A 
- \hat{g}_{L,0}^R \hat{t}_{0n}^R \hat{g}_{R,n}^K \hat{t}_{mn}^{R\dagger} \right. \right. \nonumber \\
& & \hspace{2.1cm} \left. \left. + \hat{g}_{R,0}^R \hat{t}_{n0}^{A\dagger} \hat{g}_{L,n}^K \hat{t}_{nm}^A 
- \hat{t}_{n0}^{A\dagger} \hat{g}_{L,n}^K \hat{t}_{nm}^A \hat{g}_{R,m}^A \right) \right].
\end{eqnarray}
\rrule

In order to illustrate this current formula we now investigate the dc component for a spin singlet 
superconductor with no spin active interface in two limiting cases: tunnel regime and perfect transmission. 
For the description of a poorly transmissive barrier one can solve perturbatively
the equation for $\hat{t}^{R,A}_{nm}$. Up to first order in the coupling
element: $\hat{t}^{R,A}_{LR,nm} \approx (v/2) (\hat{1} + \hat{\tau}_3) \delta_{m,n+1} \;
+ \; (v/2) (\hat{1} - \hat{\tau}_3) \delta_{m,n-1}$ and higher harmonics
can be neglected. Substituting this perturbative solution into the expression of
the dc current, $j_0$, one finds the traditional result

\begin{eqnarray}
j_0(\pfhv) & = & {\cal{D}} \int^{\infty}_{-\infty} d\epsilon \; 
\rho_L (\epsilon-eV) \rho_R (\epsilon) \times \nonumber \\
& & \left[ f_{FD}(\epsilon-eV) - f_{FD}(\epsilon) \right] ,
\label{tunnel-formula}
\end{eqnarray}

\noindent
where $f_{FD}$ is the Fermi distribution function and $\rho_{L,R}$ are
the local densities of states at the corresponding side of the interface:
$\rho_i(\epsilon) = \frac{1}{\pi} \mbox{Im} \left\{ g^A_{i}(\epsilon) \right\}$. 

In the case of a perfect transmissive contact (${\cal{D}}=1$), the absence of backscattering
greatly simplify the calculation of the different harmonics of the T-matrix
components (see Ref. [\onlinecite{Cuevas1996}]). In the case of a symmetric contact
the dc current for perfect transmission can be written as \cite{note1}

\begin{eqnarray}
j_0(\pfhv) & = & \int^{\infty}_{-\infty} d\epsilon \;
\sum^{\infty}_{m=0} \left[ \prod^{m}_{j=1} |\gamma_j|^2 \right]
\left( 1 - |\gamma_0|^2 |\gamma_{m+1}|^2 \right) \times \nonumber \\
& & \left\{ f_{FD}(\epsilon) - f_{FD}(\epsilon+(m+1)eV) \right\} ,
\end{eqnarray}

\noindent
where $\gamma_j \equiv \gamma(\epsilon+jeV)$ and $\gamma(\epsilon)$ is the
amplitude of an Andreev reflection in a N-S interface with perfect transmission.
This amplitude is defined as $\gamma^R \equiv f^R/(g^R -i\pi)$, where the Green's functions are 
evaluated at the interface. This is a very appealing formula that tell us that
the probability of a multiple Andreev reflection is basically the product of
the different Andreev reflections that take place at each side of the interface.

In order to illustrate our approach in the case of a voltage biased contact, we
consider here a junction between two d-wave superconductors. Let us analyze in particular
the symmetric junction mentioned in section IIIB, whose description is based on the 
Green's functions of Eq. (\ref{dwaveg}). Again, we neglect pair breaking effect and assume
constant order parameter up to the interface. The proper self-consistent treatment of these 
junctions will be the subject of a forthcoming publication. In this contact geometry the 
zero-energy bound states in each side of the interface strongly affect the transport
through this system. Of course, the results for the current depend on the contact model used.
Let us first consider the case of a disordered contact. In this case, the propagators which
enter in the current formula are the angle averaged ones. This implies that 
the anomalous propagators vanish, which means that the current is only due to 
single-quasiparticle processes. Thus, the current formula reduces to

\begin{equation}
j_0(V) = \int^{\infty}_{-\infty} d\epsilon \; {\cal{T}}(\epsilon,V) \left[f_{FD}(\epsilon-eV) -
f_{FD}(\epsilon) \right],
\label{disorder-formula}
\end{equation}

\noindent
where ${\cal{T}}(\epsilon,V)$ is an energy and voltage dependent transmission coefficient
given by

\begin{equation}
{\cal{T}}(\epsilon,V) = \frac{4 \pi^2 |v|^2 \langle \rho_L(\epsilon-eV) \rangle_{\pfhv}
\langle \rho_R(\epsilon) \rangle_{\pfhv}} {| 1 - |v|^2 \langle g_L(\epsilon-eV) \rangle_{\pfhv}
\langle g_R(\epsilon) \rangle_{\pfhv} |^2} ,
\label{energy-transmission}
\end{equation}

\noindent
where $\langle \rho_i(\epsilon) \rangle_{\pfhv}$ ($i=L,R$) is the local density of states at the 
interface. 

\begin{figure}
\centerline{{\epsfxsize=0.45\textwidth{\epsfbox{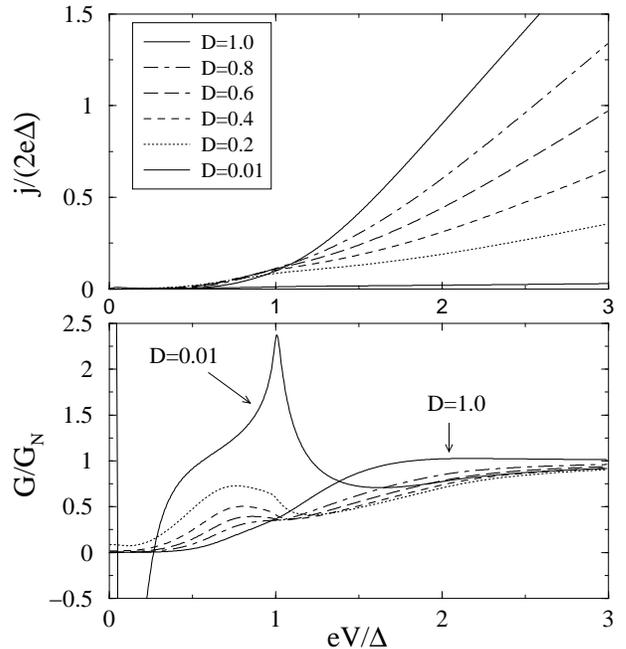}}}}
\caption{Zero temperature I-V characteristics (upper panel) and differential conductance for different
transmissions (lower panel) of a disordered contact between two d-wave superconductors. The misorientations
are $\alpha=\pi/4$. The conductance is normalized by the normal state conductance and the voltage is expressed
in units of the maximum gap $\Delta$.}
\label{av-IV}
\end{figure}

In Fig. \ref{av-IV} we show the current-voltage characteristics and differential conductance
for different values of the normal transmission coefficient ${\cal D}$ for this disordered model. The only abrupt
feature exhibited by the current inside the gap occurs in the tunnel regime. The resonant tunneling through
the zero-energy bound states leads to a zero-bias anomaly and the subsequent negative differential
conductance. As it is well known, the position and the height of the peak in the conductance depends on
intrinsic width of zero-energy states \cite{Hurd97,Barash97,Samanta98,Lofwander99}. It is known theoretically that
the elastic scattering with bulk impurities \cite{Poenicke99} or a diffusive surface layer \cite{Fogelstrom-preprints}
provide an intrinsic broadening, which for the case of Born scatteres is $\propto \sqrt{\Gamma \Delta}$,
where $\Gamma=1/2\tau$ is the effective pair-breaking parameter locally at the surface. In Fig. \ref{av-IV} we 
have introduced a small phenomenological broadening of $10^{-2} \Delta$ to mimic this intrinsic effect. 

\begin{figure}
\centerline{{\epsfxsize=0.5\textwidth{\epsfbox{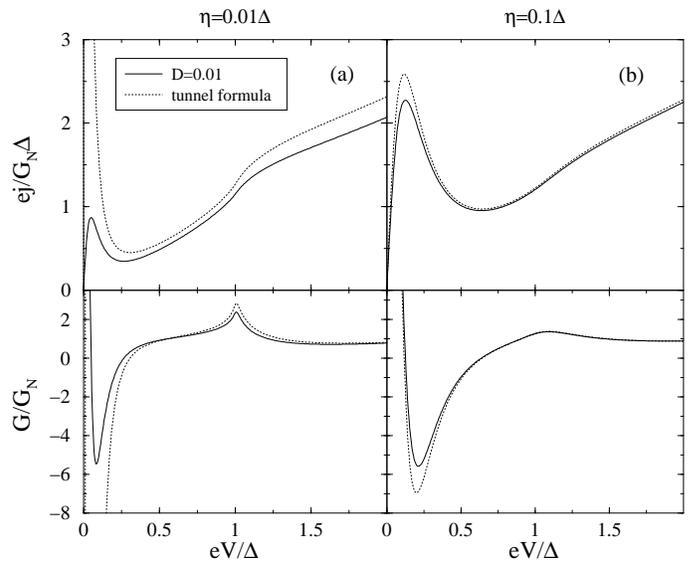}}}}
\caption{Comparison between the tunnel formula (Eq. \ref{tunnel-formula}) and the exact result of Eq. 
(\ref{disorder-formula}) for a transmission ${\cal D}=0.01$ and two different values of the broadening $\eta$: 
(a) $\eta=0.01\Delta$, and (b) $\eta=0.1\Delta$. The upper panels show the zero-temperature I-V curves and 
the lower ones the corresponding differential conductances.}
\label{tunnel}
\end{figure}

With this simple calculation we can already point out the following. Many calculations of the I-V curves 
in junctions of unconventional superconductors are limited to the tunneling regime, and make use
of the traditional tunnel formula (see Eq. \ref{tunnel-formula}). However, if bound states or divergencies
are present in the density of states, this perturbative expression does not even give the correct result for 
a low transparent contact, and one needs to include high order processes as in Eq. 
(\ref{disorder-formula}). Indeed, this is a well known problem in the context of s-wave 
superconductors (for a discussion of this problem see for instance Ref. [\onlinecite{Cuevas1996}]). In order to 
illustrate this fact, in Fig. \ref{tunnel} we compare the result of Eq. (\ref{disorder-formula}) and the 
tunneling result for a small transparency ${\cal D}=0.01$. One can see in Fig. \ref{tunnel}(a) that there is a 
clear difference between these two results. In particular, in the low voltage regime the complete expression 
renormalizes the unphysical divergencies obtained with the tunnel formula. As the broadening of the level is 
increased the difference between these two results diminishes progressively (see Fig. \ref{tunnel}(b)). Indeed, 
recent experiments in $ab$-plane tunneling into YBCO \cite{Covington1997,Aprili1998,Greene2000,Covington2000} 
suggest that the broadening of the zero bias anomaly can be as large as $25\%$ of the 
gap $\Delta$ (see Ref. [\onlinecite{Poenicke00}]). In this case, there is no significant difference 
in the resulting current between both expressions.

\begin{figure}
\centerline{{\epsfxsize=0.45\textwidth{\epsfbox{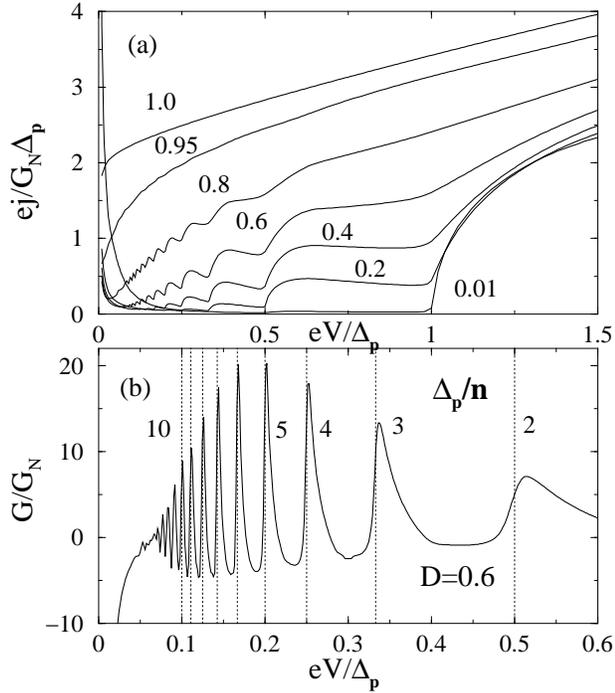}}}}
\caption{(a) Trajectory resolved zero-temperature I-V curves for the d-wave contact with a momentum conserving
interface. The different curves correspond to different values of normal transmission coefficient. In order to see
these curves in the same scale the current is normalized by the normal state conductance $G_N$ which includes
the transmission coefficient. Moreover, the current and the voltage are normalized in units of the momentum-dependent 
gap. (b) Differential conductance for a transmission ${\cal D}=0.6$. The vertical lines indicate the 
positions $eV_n=\Delta_p/n$, $n=2,...,10$, as a guide for the eye.}
\label{IV-dwave}
\end{figure}

Let us now consider the case of a momentum conserving contact, which is the usual assumption
in the Zaitsev boundary conditions. To gain some insight into the final result, in Fig. \ref{IV-dwave}
we show the contribution of an individual trajectory $\pfhv$. The current and voltage are normalized in
units of the gap seen by this trajectory. As can be observed, the current exhibits a pronounced
subgap structure at voltages $eV=\Delta_p /n$, where $n$ is a integer number, together with the 
appearance of negative differential conductance (this can be seen better in the lower panel of this figure).
These features are a simple consequence of the resonant tunneling across the zero-energy bound states. Indeed, 
this type of I-V has been previously obtained in the context of a junction between two conventional
superconductors coupled by means of a resonant transmission (see Ref. [\onlinecite{Levy97,Johansson99}]).
Notice also the presence of a zero bias peak, specially clear for low transparencies, and which is
a consequence of the small broadening introduced in the calculation.
The total current is obtained by averaging over the different trajectories. Thus, the final result depends on the
model for the angular dependence of the normal transmission coefficient. With any reasonable
reasonable model most of the features of the trajectory resolved current disappear. In particular,
the subharmonic gap structure is washed out, and it only remains a peak in the conductance at $eV \approx 
\Delta$ (see for instance Ref. [\onlinecite{Hurd97}]).

\section{Current fluctuations}

During the last years it has become progressively clearer that a deep
understanding of the electronic transport in mesoscopic systems requires
the analysis of quantities which goes beyond the straightforward measurement
of the current-voltage characteristics. In this sense the noise or
time-dependent current fluctuations has emerged as a very useful tool which
provides new information on the time correlations of the current,
information about channel distributions, statistics and charge of the
carriers \cite{Noise}. In the case of superconducting contacts, most of the activity
has been restricted to the case of s-wave superconductors \cite{Averin96,Dieleman97,Cuevas99,Naveh99}. 
In the case of unconventional superconductors there are only a few theoretical works in the context of 
hybrid structures like normal metal/d-wave superconductors \cite{Zhu99,Tanaka00}. We believe that
in the next future the measurement of current fluctuations will be an important tool for a deeper
understanding of the symmetry of the order parameter and origin of the superconductivity in general
in the case of HTS materials. For this reason in this section we describe the calculation of the
noise spectrum within our approach, valid for any type of contact between unconventional superconductors.

Let us remember that the noise is characterized by its spectral density or
power spectrum $S(\omega)$, which is simply the Fourier transform at frequency
$\omega$ of the current-current correlation function

\begin{eqnarray}
S(\omega) & = & \int d (t^{\prime}-t) \; e^{i\omega \left(t^{\prime}-t\right)} \langle \delta 
\hat{\j}(t^{\prime}) \delta \hat{\j}(t) + \delta \hat{\j}(t) \delta\hat{\j}(t^{\prime})
\rangle \nonumber \\
& \equiv & \int d (t^{\prime}-t) \; e^{i\omega \left(t^{\prime}-t\right)} \; K(t,t^{\prime}),
\end{eqnarray}

\noindent
where $\delta \hat{\j}(t)= \hat{\j}(t) - <\hat{\j}(t)>$ are the fluctuations in the current.

In order to obtain the expression of the current-current correlator, first we need
an expression for the current operator. Within our model this operator evaluated at the interface
can written as follows

\begin{equation}
\hat{\j}(t) = ie \sum_{\sigma} \left\{ v_{LR,\sigma} \hat{c}^{\dagger}_{L,\sigma}(t) 
\hat{c}_{R,\sigma}(t) \; - \; v_{RL,\sigma} \hat{c}^{\dagger}_{R,\sigma}(t) \hat{c}_{L,\sigma}(t) \right\}.
\end{equation}

\noindent
This expression is a simple consequence of the continuity equation for the current \cite{Caroli71}.

In order to calculate the noise we need in principle to evaluate correlators
of four field operators. However, we are working in the framework of a mean
field theory, which means that we can decouple these correlators in
terms of one-particle Green's functions using the Wick theorem. With this in mind, it is straightforward
to show that the kernel $K(t,t^{\prime})$ can be expressed in terms
of the interface Keldysh Green's functions as follows \cite{Cuevas99}

\begin{eqnarray}
K(t,t^{\prime}) = e^2 \left\{ \mbox{Tr} \left[
\hat{v}_{RL} \hat{G}^{<}_{LL}(t,t^{\prime}) \hat{v}_{LR} \hat{G}^{>}_{RR}(t^{\prime},t)
\right. \right. & & \nonumber \\
\hspace*{-2cm}
+ \hat{v}_{LR} \hat{G}^{<}_{RR}(t,t^{\prime}) \hat{v}_{RL} \hat{G}^{>}_{LL}(t^{\prime},t) & &
\nonumber \\ \hspace{-2cm}
- \hat{v}_{RL} \hat{G}^{<}_{LR}(t,t^{\prime}) \hat{v}_{RL} \hat{G}^{>}_{LR}(t^{\prime},t) & &
\nonumber \\ \hspace{-2cm} \left. \left.
- \hat{v}_{LR} \hat{G}^{<}_{RL}(t,t^{\prime}) \hat{v}_{LR} \hat{G}^{>}_{RL}(t^{\prime},t) \right]
+ (t \rightarrow t^{\prime}) \right\} & & ,
\end{eqnarray}

\noindent
where the functions $\hat{G}^{<}$ and $\hat{G}^{>}$ are related to the usual
advanced, retarded and Keldysh functions in the following way

\begin{eqnarray}
\hat{G}^< & = & \left( \hat{G}^K - \hat{G}^R + \hat{G}^A \right)/2
\nonumber \\
\hat{G}^> & = & \left( \hat{G}^K + \hat{G}^R - \hat{G}^A \right)/2.
\end{eqnarray}

In order to compactify the notation, we introduce the trace $\tilde{\mbox{Tr}}$ and the matrix $\tilde{\tau}_3$
which act in the ``reservoir'' space. Then, the noise kernel reads

\begin{eqnarray}
K(t,t^{\prime}) & = & -e^2 \tilde{\mbox{Tr}} \left[
\tilde{v} \tilde{G}^{<}(t,t^{\prime}) \tilde{\tau}_3 \tilde{v} \tilde{G}^{>}(t^{\prime},t) 
+ \right. \nonumber \\ & & \hspace{1.2cm} \left.
\tilde{v} \tilde{G}^{>}(t,t^{\prime}) \tilde{\tau}_3 \tilde{v} \tilde{G}^{<}(t^{\prime},t)
\right].
\end{eqnarray}

Now, in order to eliminate the Green's functions in favor of the T-matrix elements, we use the relation

\begin{equation}
\tilde{G}^{<,>} = \left( \tilde{1} + \tilde{G}^R \circ \tilde{v} \right) \circ \tilde{G}^{<,>}_{\infty} \circ
\left( \tilde{1} + \tilde{v} \circ \tilde{G}^A \right),
\end{equation}

\noindent
where $\tilde{G}^{<}_{\infty}(\epsilon) = \left[ \tilde{G}^A_{\infty}(\epsilon) -
\tilde{G}^R_{\infty}(\epsilon) \right] f_{FD}(\epsilon)$ and $\tilde{G}^{>}_{\infty}(\epsilon) = 
\left[ \tilde{G}^A_{\infty}(\epsilon) - \tilde{G}^R_{\infty}(\epsilon) \right] 
\left( f_{FD}(\epsilon) - 1 \right)$. Now, making use of Eqs. (3-7) it is easy to see that the
following relation holds

\begin{equation}
\tilde{v} \circ \tilde{G}^{<,>} = \tilde{T}^R \circ \tilde{G}^{<,>}_{\infty} \circ \left( \tilde{1} + \tilde{T}^A 
\circ \tilde{G}^A_{\infty} \right).
\end{equation}

\lrule
\noindent
This expression allows us to write the noise kernel as follows

\begin{equation}
K(t,t^{\prime})  =  -e^2 \tilde{\mbox{Tr}} \left\{
\left[ \tilde{T}^R \circ \tilde{G}^{<}_{\infty} \circ \left( \tilde{1} + \tilde{T}^A \circ \tilde{G}^A_{\infty} 
\right) \right] (t,t^{\prime}) \; \tilde{\tau}_3 \; 
\left[ \tilde{T}^R \circ \tilde{G}^{>}_{\infty} \circ \left( \tilde{1} + \tilde{T}^A \circ \tilde{G}^A_{\infty} 
\right) \right] (t^{\prime},t) + (t \rightarrow t^{\prime}) \right\}.
\end{equation}

%\begin{eqnarray}
%K(t,t^{\prime}) & = & -e^2 \tilde{\mbox{Tr}} \left\{
%\left[ \tilde{T}^R \circ \tilde{G}^{<}_{\infty} \circ \left( \tilde{1} + \tilde{T}^A \circ \tilde{G}^A_{\infty} 
%\right) \right] (t,t^{\prime}) \; \tilde{\tau}_3 \times \right. \nonumber \\ & & \hspace{-1cm} \left.
%\left[ \tilde{T}^R \circ \tilde{G}^{>}_{\infty} \circ \left( \tilde{1} + \tilde{T}^A \circ \tilde{G}^A_{\infty} 
%\right) \right] (t^{\prime},t) + (t \rightarrow t^{\prime}) \right\}.
%\end{eqnarray}

As explained in section II, once we have eliminated the full Green's functions in the noise kernel,
we can perform the quasiclassical $\xi$-integration and substitute the quasiclassical Green's functions,
$\tilde{g}_{\infty}$, for the full ones, $\tilde{G}_{\infty}$. Thus, the noise kernel can be expressed
finally as

\begin{equation}
K(t,t^{\prime})  =  -e^2 \tilde{\mbox{Tr}} \left\{ \left[ 
\tilde{T}^R \otimes \tilde{g}^{<}_{\infty} \otimes \left( \tilde{1} + \tilde{T}^A \otimes \tilde{g}^A_{\infty} 
\right) \right] (t,t^{\prime}) \; \tilde{\tau}_3 \; 
\left[ \tilde{T}^R \otimes \tilde{g}^{>}_{\infty} \otimes \left( \tilde{1} + \tilde{T}^A \otimes 
\tilde{g}^A_{\infty} \right) \right] (t^{\prime},t) + (t \rightarrow t^{\prime}) \right\}.
\end{equation}

%\begin{eqnarray}
%K(t,t^{\prime}) & = & -e^2 \tilde{\mbox{Tr}} \left\{ \left[ 
%\tilde{T}^R \otimes \tilde{g}^{<}_{\infty} \otimes \left( \tilde{1} + \tilde{T}^A \otimes \tilde{g}^A_{\infty} 
%\right) \right] (t,t^{\prime}) \; \tilde{\tau}_3 \times \right. \nonumber \\ & & \hspace{-1cm} \left.
%\left[ \tilde{T}^R \otimes \tilde{g}^{>}_{\infty} \otimes \left( \tilde{1} + \tilde{T}^A \otimes 
%\tilde{g}^A_{\infty} \right) \right] (t^{\prime},t) + (t \rightarrow t^{\prime}) \right\}.
%\end{eqnarray}

Let us stick now to the case of a constant bias voltage applied across the interface. In this case, for both
contact models considered in section III, we can resolve the current fluctuations in trajectories as follows

\begin{equation}
S(\omega,t) = e^2 N_F \langle S(\pfhv,\omega,t) \rangle_{\pfhv},
\end{equation}
\noindent
where the time-dependent contribution of given trajectory with momentum $\pfhv$ can be written as

\begin{equation}
S(\pfhv,\omega,t) = \sum_{m=-\infty}^{\infty} S_m(\pfhv,\omega) e^{im \phi(t)} ,
\end{equation}
\noindent
where the different ac-components of the noise can be expressed in terms of the Fourier component
of the T-matrix elements in the following way

\begin{equation}
S_m(\pfhv,\omega)  =  - \int d\epsilon \sum_n \tilde{\mbox{Tr}} \left\{ \left[ 
\tilde{T}^R \tilde{g}^{<}_{\infty} \left( \tilde{1} + \tilde{T}^A \tilde{g}^A_{\infty} \right) \right]_{0n}
(\epsilon) \; \tilde{\tau}_3 \; 
\left[ \tilde{T}^R \tilde{g}^{>}_{\infty} \left( \tilde{1} + \tilde{T}^A \tilde{g}^A_{\infty} \right) 
\right]_{nm} (\epsilon + \omega) + (\epsilon \rightarrow \epsilon + \omega) \right\}.
\end{equation}
\rrule

%\begin{eqnarray}
%S_m(\pfhv,\omega) & = & -\frac{e^2}{\hbar^2} \int d\epsilon \sum_n \tilde{\mbox{Tr}} \left\{ \left[ 
%\tilde{T}^R \tilde{g}^{<}_{\infty} \left( \tilde{1} + \tilde{T}^A \tilde{g}^A_{\infty} \right) \right]_{0n}
%(\epsilon) \times \right. \nonumber \\ & & \hspace{-2cm} \tilde{\tau}_3 \; \left.
%\left[ \tilde{T}^R \tilde{g}^{>}_{\infty} \left( \tilde{1} + \tilde{T}^A \tilde{g}^A_{\infty} \right) 
%\right]_{nm} (\epsilon + \omega) + (\epsilon \rightarrow \epsilon + \omega) \right\}.
%\end{eqnarray}

Notice that in the case of a junction between two superconductors, the noise, as the current, oscillates on 
time with all the harmonics of the Josephson frequency. Notice also that we have reduced the calculation of 
this quantity to the determination of the different Fourier components of the T-matrix elements, which has been 
detailed in section IV.

In order to illustrate the calculation of the current fluctuations, we consider the contact between
d-wave superconductors analyzed in the previous section. In particular, we present results for the
zero-frequency noise at zero temperature, $S$, i.e. the zero-frequency shot noise. At this point, it is worth
remarking that by zero-frequency noise we mean noise at a frequency lower than any relevant energy scale in
our problem, gap for instance, and high enough to neglect $1/f$ noise \cite{Koelle99}. Again, the final result
depends on the type of contact model under investigation. Let us start discussing the shot noise for the
disordered contact. In this case, due to the vanishing of the anomalous Green's functions, the whole calculation 
reduces to the determination of the quasiparticle contribution, which in term of the transmission coefficient 
of Eq. (\ref{energy-transmission}) can be written as

\begin{equation}
S = \int^{eV}_{0} d\epsilon \; {\cal{T}}(\epsilon,V) \left( 1 - {\cal{T}}(\epsilon,V) \right).
\label{noise-disorder}
\end{equation}

\begin{figure}
\centerline{{\epsfxsize=0.45\textwidth{\epsfbox{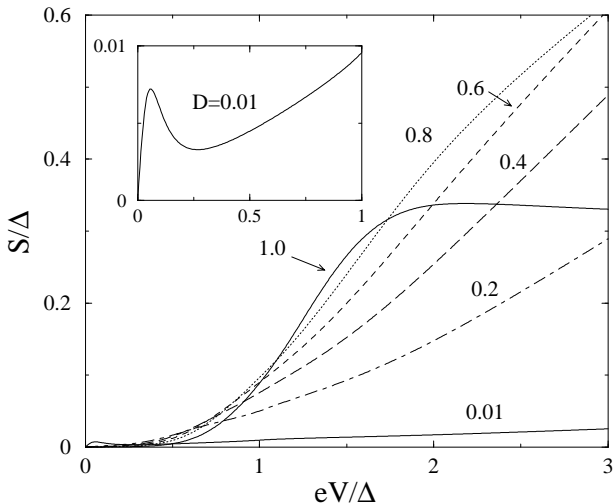}}}}
\caption{Zero-frequency shot noise for the disordered contact considered in Fig. \ref{tunnel}. The inset show the
low bias limit of the curve in the tunneling regime (${\cal D}=0.01$).}
\label{noise1}
\end{figure}

\noindent
This is simply the result that one obtains for a normal contact with an energy and voltage dependent transmission
coefficient \cite{Noise}. In Fig. \ref{noise1} we show the result of Eq. (\ref{noise-disorder}) for different normal 
transmissions. In the tunneling regime the shot noise is $S(V) \approx 2ej(V)$ and the most remarkable feature is 
the zero bias anomaly (see inset of Fig. \ref{noise1}). In the case of perfect transmission there is a non-zero 
noise due to the fact that the transmission coefficient ${\cal{T}}(\epsilon,V)$ is less than one in the gap region. 
For voltages much greater than the maximum gap, then ${\cal{T}}(\epsilon \gg \Delta, {\cal D}=1.0) \rightarrow 1$, 
what makes that the noise at ${\cal D}=1.0$ saturates in the high voltage regime.

\begin{figure}
\centerline{{\epsfxsize=0.5\textwidth{\epsfbox{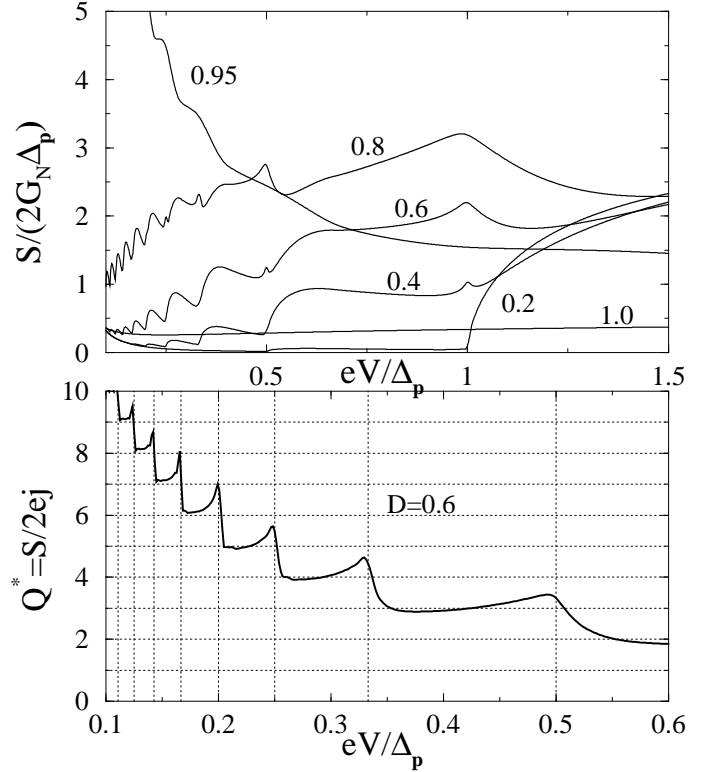}}}}
\caption{The upper panel shows the trajectory resolved zero-frequency shot noise for the momentum conserving 
case considered in Fig. \ref{IV-dwave}. The shot noise and voltage are normalized by the trajectory-dependent gap 
$\Delta_p$. In the lower panel one can see the effective charge defined as $Q^*=S/2ej$ as a
function of voltage for a transmission ${\cal D}=0.6$. The vertical lines indicate the position of the voltages
$eV_n=\Delta_p/n$, $n=2,...,10$.}
\label{noise2}
\end{figure}

More interesting is the case of the momentum conserving interface. In Fig. \ref{noise2} we show the contribution
of a trajectory of momentum $\pfhv$. As in the case of the current, the shot noise exhibits a rich subharmonic
gap structure, which persists almost up to perfect transmission. The shape of the different curves can
be understood in the same terms as the BCS case (see Ref. [\onlinecite{Cuevas99}]), with the additional ingredient
of the resonant tunneling through the zero energy states. Of course, as in the case of the current, most of
these features disappear after performing the angular average.

In the case of conventional superconductors, the shot noise has been proposed as a tool for measuring the 
multiple charge quanta transferred by the multiple Andrev reflections \cite{Cuevas99}. Obviously we can pose 
here the same question in the case of unconventional superconductors. Indeed, a noise experiment has been 
recently proposed by Auerbach and Altman \cite{Auerbach00} to discriminate between two possible explanations
of the pronounced subharmonic gap structure observed in YBCO edge junctions \cite{Nesher99}. Namely, usual multiple 
Andreev reflections in a d-wave superconductor and magnon pair creation in the context of the SO(5) theory. In this 
latter case the observed charge should be $Q^*=2ne$, where $n=1,2,...$, at a voltage $eV_n=\Delta/n$. This result 
has to be compared with $Q^*=ne$ expected in the traditional view of MAR. In order to contribute to the solution of 
this puzzle, we show in Fig. \ref{noise2} (lower panel) the effective charge, $Q^*=S/2ej$, for a transmission 
${\cal D}=0.6$. This result confirm the traditional interpretation that in the MAR process of order $n$ a charge 
$ne$ is transferred. Usually, in order to observe a clear quantization of the charge one should go to the tunneling 
regime \cite{Cuevas99}, but in this case this is not necessary due to the resonant tunneling through the zero-energy 
states. Notice again that this is the contribution of a single trajectory and after angle averaging this clear 
quantization of the charge with voltage disappears. The exhaustive analysis of the shot noise in d-wave contacts
will be presented in a forthcoming publication.

%===============================================================================================

\section{Conclusions}

We have shown how a Hamiltonian approach and the quasiclassical theory of superconductivity
can be combined to give a powerful tool to analyze electronic and transport properties
of superconducting junctions. In particular, we have demonstrated that a simple Hamiltonian description
of an interface can be used to model a great variety of contacts. This
Hamiltonian description can be brought into quasiclassical theory via a T-matrix equation, resulting in
a new formulation of boundary conditions. These boundary conditions do not contain any spurious solution
and can be efficiently solved to compute any transport property. The broad applicability of this 
formulation covers cases ranging from conventional superconductors to unconventional ones, clean systems 
and diffusive ones. Moreover, it can be applied to spin active interfaces and it is well suited for the 
description of time-dependent phenomena like the I-V characteristics and the noise properties of junctions 
with arbitrary 
transmission and bias voltage. We have illustrated this approach with the calculation of Josephson current in 
a great variety of situations. The calculation of I-V characteristics and the noise has been exemplified
with the analysis of a contact between two d-wave superconductors. In particular, we have briefly discussed
the use of shot noise as a possible tool for measuring the charge of the Andreev reflections in 
unconventional superconductors.

\acknowledgements
We thank A. Poenicke, A. Mart\'{\i}n-Rodero, A. Levy Yeyati and G. Sch\"on for fruitful discussions. 
J.C. Cuevas acknowledges the European Community for the Marie Curie Fellowship under contract HPMF-CT-1999-00165. 
M.~Fogelstr\"om thanks the Swedish Natural Science Research Council (NFR) under contract F620-1072/2000
and the DFG project SFB 195 for support.

\end{multicols}

\begin{references}
\bibitem{note1} 
The term $m=0$ in the series appearing in the integrand should be understood as follows: 
$(1- |\gamma_0|^2 |\gamma_1|^2) \left\{ f_{FD}(\epsilon) - f_{FD}(\epsilon+eV) \right\}$.
\end{references}
\end{document}